\documentclass[11pt]{article}
\pdfoutput=1 
\usepackage{nicefrac} 
\usepackage{jheppubg} 
\usepackage[T1]{fontenc} 
\usepackage{accents} 
\newcommand{\beq}{\begin{equation}}
\newcommand{\eeq}{\end{equation}}
\newcommand{\beqa}{\begin{eqnarray*}}
\newcommand{\eeqa}{\end{eqnarray*}}
\def \al{\alpha}
\def \be{\beta}
\def \ga{\gamma}
\def \de{\delta}
\def \ep{\epsilon}

\def \et{\eta}

\def \rh{\rho}
\def \si{\sigma}
\def \ta{\tau}

\def \ph{\phi}
\def \ch{\chi}

\def \om{\omega}
\def \Ga{\Gamma}
\def \De{\Delta}
\def \Th{\Theta}
\def \La{\Lambda}

\def \Ph{\Phi}
\def \Ps{\Psi}
\def \Om{\Omega}
\def \pa{\partial}

\def \lb{\left[}
\def \rb{\right]}
\def \lp{\left(}
\def \rp{\right)}

\def \ll{\left|}
\def \rl{\right|}

\def \p#1{\phantom{#1}}

\def \fr#1#2{{\textstyle \frac{#1}{#2}}}
\def \ha{\fr{1}{2}}

\def \rbar{\smash{\ensuremath \raisebox{.2pt}{$\scriptscriptstyle\_$}}}
\def \f#1{\underaccent{\rbar}{{#1}}}
\def \rbard{\smash{\ensuremath \raisebox{1.8pt}{$\scriptscriptstyle\_$}}}
\def \fd#1{\underaccent{\rbard}{{#1}}}
\def \ff#1{\underaccent{=}{{#1}}}

\def \nf#1{\underaccent{\!\p{.} \sim}{{#1}}}
\def \rhup{\smash{\ensuremath \raisebox{-1.5pt}{$\scriptscriptstyle\rightharpoonup$}}}
\def \ve#1{\accentset{\rhup}{{#1}}}

\def \ud#1{\underaccent{\,.}{{#1}}}
\def \udd#1{\underaccent{..}{{#1}}}
\def \dd{\smash{\ensuremath \raisebox{-3.0pt}{$\scriptscriptstyle{- \!\p{.}\! \smash{\cdot}}$}}}
\def \udf#1{\underaccent{\dd}{{#1}}}
\def \udff#1{\underaccent{= \!\p{.}\! \smash{\cdot}}{{#1}}}
\def \od#1{\accentset{\cdot}{{#1}}}

\DeclareMathOperator{\Tr}{Tr}
\def \arXiv#1{\texttt{\href{http://arxiv.org/abs/#1}{arXiv:#1}}}

\title{\boldmath Lie Group Cosmology}

\author{A. Garrett Lisi}

\affiliation{Pacific Science Institute, Makawao, HI, USA}

\emailAdd{Gar@Li.si}

\abstract{Our universe is a deforming Lie group.}

\begin{document} 

\maketitle

\vspace{30em}

\section{Introduction}

To the best of our current knowledge, the universe exists as quantized excitations of a fiber bundle. The base space of this fiber bundle is our four-dimensional spacetime manifold, while the fibers over spacetime are Lie group manifolds and representation spaces. Elementary particles are quantized excitations of three kinds of fields over spacetime: $1$-forms, scalars, and fermions (anti-commuting Grassmann number fields), each valued in their corresponding fiber. Among these fields, the spacetime frame $1$-form determines spatial lengths and temporal durations in spacetime and is valued in the vector representation space of the $Spin(1,3)$ Lie group of the gravitational spin connection. The curvature of this spin connection is the pseudo-Riemannian spacetime curvature, with dynamics governed by General Relativity. The Higgs scalar field and all fermions are valued in fundamental representation spaces of the Standard Model Lie group, $U(1) \times SU(2) \times SU(3) \, / \, \mathbb{Z}_6$, with the fermions also spinors under $Spin(1,3)$. The spin connection and Standard Model gauge fields correspond to connection $1$-forms over their respective principal fiber bundles and necessitate the existence of anti-commuting BRST ghost fields valued in the same Lie algebra. Considered as a whole, our universe is the \emph{total space} manifold consisting of base spacetime and all these fibers together, with field configurations corresponding to sections of and over this bundle, its geometry described by curvature, and dynamics prescribed by Quantum Field Theory.

Why? Why do precisely these structures exist? Why this specific interplay of base manifold, frame, Lie groups, Grassmann numbers, scalars, representation space fibers, sections, connections, and ghosts? Is our universe fundamentally a mess, or is there some simple and natural structure that all this could emerge from, or be parts of? One approach to answering these questions is String Theory (or, more generally, M-Theory), but string unification models have grown excessively in complexity while producing zero predictive progress. After several decades of extensive theoretical exploration leading nowhere, it is time to consider that the string program may have been a wrong turn. If we backtrack, imagining String Theory never happened, we can go in a new direction, building on the success of Grand Unified Theories and recent progress in Loop Quantum Gravity. The structures of GUTs and LQG rely heavily on Lie groups and are remarkably compatible. By considering the known Lie groups and fields of physics as parts of a larger geometric whole, we move towards Lie group unification.

Grand Unified Theories are elegant and successful. They embed the Standard Model Lie group, $U(1) \times SU(2) \times SU(3) \, / \, \mathbb{Z}_6$, inside larger groups, such as $Spin(10)$, with Higgs and fermion multiplets in various representation spaces~\cite{Baez}. The principal geometric structure of a unified gauge theory is an Ehresmann connection, a Lie algebra valued $1$-form over the total space of the fiber bundle, with the Yang-Mills action (including a metric-dependent Hodge dual) and quantum description following from renormalizable Quantum Field Theory, and spacetime corresponding to the sheaf of gauge-related sections over a four-dimensional base. In a similar spirit, MacDowell and Mansouri~\cite{MM} proposed a description of gravity as a $Spin(1,4)$ gauge theory, with the $Spin(1,4)$ connection (over a fourteen-dimensional total space) breaking into a $Spin(1,3)$ connection and a gravitational frame over a four-dimensional base. (The gravitational spin connection is the fundamental field in most modern approaches to quantum gravity.) Although the MacDowell-Mansouri model is algebraically consistent, it has several problems. From a viewpoint of efficiency, the fourteen dimensions of the total space seems unnecessarily large for a description of gravity. Also, it is awkward to describe matter within the MacDowell-Mansouri framework, which does not naturally have a Hodge duality operator. Crucially, in the MacDowell-Mansouri model and related gauge theories, there is no reason why part of the $Spin(1,4)$ connection should relate to spatial lengths and temporal durations in the base space. There is no natural direct relationship between the frame part of the Lie algebra and motion in the base manifold.

The ontological problems with the MacDowell-Mansouri model are challenging, but there is a structurally different geometric picture in which all of these problems are solved. By allowing the ten-dimensional $Spin(1,4)$ Lie group to \emph{deform} while maintaining the rigidity of its $Spin(1,3)$ subgroup, we get a succinct description of gravity as \emph{Cartan geometry}~\cite{Sharpe, Wise, Hans}. The four-dimensional coset representative of the coset $Spin(1,4)/Spin(1,3)$ embedded in $Spin(1,4)$ may be identified as \emph{de Sitter spacetime} (half of de Sitter space, $dS/\mathbb{Z}_2$). In the deforming (or ``excited'') Lie group, $S\tilde{pin}(1,4)$, the embedded four-dimensional coset representative subspace, and the sheaf of equivalent subspaces related by $Spin(1,3)$ transformations, are allowed to become wavy. This is described by $Spin(1,4)$'s natural Maurer-Cartan form deforming and becoming a \emph{Cartan connection}, consisting of the $Spin(1,3)$ connection and gravitational frame for the embedded wavy spacetime. In this Cartan geometry, the ten-dimensional deforming Lie group is lower-dimensional than the fourteen-dimensional total space of the MacDowell-Mansiouri model. And, most importantly, the frame part of the $Spin(1,4)$ connection is now naturally the gravitational frame on spacetime; it is the frame on spacetime embedded in the deforming Lie group. This deforming Lie group geometry provides a Hodge duality operator, with a Yang-Mills action describing the dynamics of gravity and matter. Cartan geometry is a very efficient geometric picture, with spacetime and all its structure emerging from a single deforming Lie group.

Given the success of $Spin(10)$ Grand Unification for the Standard Model, and of Cartan geometry for gravity, it is natural to consider the further unification of gauge theory and the gravitational spin connection as parts of a $Spin(11,3)$ connection~\cite{Smolin, Percacci}. However, in the context of Cartan geometry, we immediately encounter the difficulty that the quotient space, $Spin(11,3)/(Spin(1,3) \times Spin(10)/\mathbb{Z}_2)$, of the relevant Cartan geometry is forty-dimensional---intractably large as a model for spacetime. For large Lie groups, conventional Cartan geometry produces spacetimes that are unrealistically high-dimensional. A generalization of Cartan geometry is needed.

From the perspective of Lie group deformations, we can consider what happens when a $Spin(1,4)$ subgroup of a larger Lie group, such as $Spin(12,4)$, deforms, and how that is described, with four-dimensional embedded spacetime. The \emph{generalized Cartan connection} of such a large, deforming Lie group can include precisely the gauge, spin connection, Higgs, and gravitational frame parts we need. Also, the Yang-Mills action for this generalized Cartan connection, integrated over the entire deforming Lie group, naturally reduces to a desirable action over four-dimensional spacetime.
 
With Standard Model and gravitational bosons described as parts of a generalized Cartan connection, one must wonder what is the best possible way to describe fermions. The conventional description of fermions would be to presume their existence a priori, as separate multiplets of anti-commuting (``fermionic,'' Grassmann number) fields valued in the spinor representation space of the relevant spin group. That would certainly work, but it does not seem geometrically well justified. A slightly better description of fermions is as $1$-forms on the space of connections. This approach is familiar from geometric descriptions of BRST ghosts, and is fairly promising. For that to work, with a connection valued in the Lie algebra of the relevant group, the fermions would also necessarily be valued in part of that Lie algebra, which, since they are spinors, requires that any simple unification group be an exceptional Lie group. Another possible description of fermions, even more geometric but less conventional, is to consider them as $1$-forms in the Lie group, orthogonal to embedded spacetime. In this description, the \emph{extended generalized Cartan connection} is a \emph{superconnection}, a Lie algebra valued $1$-form over the deforming Lie group, consisting of a bosonic part (spacetime $1$-form) plus a fermionic part ($1$-form orthogonal to spacetime). The most natural possible action for this geometry is the \emph{extended Yang-Mills action} for the superconnection over the entire deforming Lie group, producing a second-order action for the fermions.

If we adopt this unified description of bosons and fermions, the most natural finite-dimensional simple Lie group candidates for a complete unification are the complex, split, or quaternionic real forms of the largest simple exceptional group, $E_8$~\cite{Lisi1, Lisi2}. But there are known problems with this theory as it was previously proposed. In $E_8$ unification, once a specific $Spin(1,3)$ subgroup is chosen for gravity, there are not three generations of fermionic spinors. (This issue was discussed in the original proposal~\cite{Lisi1}, and while some consider it a terminal flaw~\cite{Distler}, we are here reframing the geometry in a way that solves this problem.) In the original proposal it was suggested that the \emph{triality} automorphism of $E_8$ might relate three different $Spin(1,3)$'s and three separate sets of fermions, each of which would be a different Standard Model generation with respect to its corresponding gravitational $Spin(1,3)$. While that suggestion did not make perfect sense within the geometric context of that $E_8$ principal bundle model, it does make sense within our extended generalized Cartan geometry, in which there are three sheaves of triality-related spacetimes embedded in deforming $\tilde{E_8}$, each with a different corresponding $Spin(1,3)$ subgroup and associated generation of fermions. With our physical spacetime a superposition of these three triality-related spacetimes, we have three generations of fermions mixing and interacting with the Higgs fields to get their masses.

The purpose of this work is not to provide a detailed derivation of all Standard Model properties, but to lay the geometric foundation of a theory from which these properties might be derived, clarifying the model proposed in~\cite{Lisi1}. In developing this minimal, unified description of all fundamental fields, we use elementary structures of differential geometry and group theory. Although the language employed is mathematically antiquated---perhaps more familiar to mathematicians of the early twentieth century than the twenty-first---it was this mathematical context from which the structures of Quantum Field Theory were born, and it is arguably still the best in which to describe them. We also rely on the philosophy of naturalness in differential geometry: Everything should be describable using primitive geometric elements, such as vector fields and forms on manifolds. Further, since our goal is to describe structures in physics usually expressed in local coordinates, we employ such descriptions here, for familiarity and concreteness. We begin with a casual description of the geometry of Lie groups, principal bundles, and General Relativity, before proceeding to a description of Cartan geometry, its generalization, and some new ideas. An advanced reader wishing to avoid pedagogy and get to the heart of the proposal may wish to skip ahead to the two-page Summary, \S\ref{sec:summary}.

\section{Lie group geometry}

A Lie group has elements, $g(x) \in G$, corresponding to points, $x$, of an $n$-dimensional manifold identified with $G$, having patches with local coordinates, $x^i$. This manifold has a specific geometric structure corresponding to the Lie group product. The right action of any group element, $h$, on all other elements of the group via the group product provides a diffeomorphism, $\ph_h : G \to G$, of the group manifold,
\beq
R_h g = g h = g(\ph_h(x)) \label{ract}
\eeq
Group elements, typically of a matrix Lie group, near the identity,
$$
h = e^{t \, v^A T_A} \simeq 1 + t \, v^A T_A
$$
may be specified by exponentiating matrix Lie algebra elements, $v=v^A T_A \in Lie(G)$, times a small parameter, $t$. The $n$ Lie algebra \emph{basis generators}, $T_A$, represented by matrices, span the $n$-dimensional vector space corresponding to small departures from the identity point. The right action of such group elements, $h$, on $G$ corresponds to a flow diffeomorphism,
$$
\ph^i_h(x) \simeq x^i + t \, v^A \xi_A{}^i(x)
$$ 
moving every point, $g(x)$, of $G$ along a vector field, $v^A \ve{\xi}_A$, determined by (\ref{ract}) for small $t$, 
$$
g + t \, v^A g \,T_A \simeq g + t \, v^A \xi_A{}^i \pa_i g(x)
$$
So, for each of the Lie algebra basis generators, $T_A$, there is a corresponding flow vector field, $\ve{\xi}_A(x)$, acting via the Lie derivative, such that\footnote{In our vector-form notation, forms are underlined according to grade and presumed wedged, and vectors act on forms to their right,
$$
\ve{v} \f{u} \f{w} = (\ve{v} \f{u}) \f{w} - \f{u} (\ve{v} \f{w}) \equiv {\bf i}_{\ve{v}} \, u \wedge w
$$}
\beq
g \, T_A = {\cal L}_{\ve{\xi}_A} g = \ve{\xi}_A \f{d} g = \xi_A{}^i \pa_i g(x) \label{liedef}
\eeq
Using this defining equation, the Lie derivative between generator vector fields is seen to be equivalent to the commutator bracket between Lie algebra generators, with the same structure constants,
\begin{eqnarray*}
\lb \ve{\xi}_A, \ve{\xi}_B \rb_{\cal L} &=& {\cal L}_{\ve{\xi}_A} \ve{\xi}_B
= \ve{\xi}_A \f{\pa} \ve{\xi}_B - \ve{\xi}_B \f{\pa} \ve{\xi}_A = C_{AB}{}^C \ve{\xi}_C \\
\lb T_A, T_B \rb &=& T_A T_B - T_B T_A = C_{AB}{}^C T_C
\end{eqnarray*}
The generator vector field component matrix, $\xi_A{}^i$, may be inverted, finding $\xi_i{}^A$ such that $\xi_A{}^i \xi_i{}^B = \de_A^B$, corresponding to a set of $1$-forms, $\fd{\xi}^A = \f{dx}^i \xi_i{}^A$. Multiplying these by the Lie algebra basis generators and using (\ref{liedef}), we find the \emph{Maurer-Cartan form} over the Lie group manifold,\footnote{We use a shorthand notation for inverses, $$g^- \equiv g^{-1}$$}
\beq
\f{\Th} = \fd{\xi}^A T_A = g^- \f{d} g \label{MC}
\eeq
The Maurer-Cartan form, considered as a Lie algebra-valued connection, encodes the local geometry of the Lie group manifold. Its curvature vanishes,\footnote{The product of $1$-forms here equates to half their Lie bracket,
$$
\f{\Th} \, \f{\Th} = \fd{\xi}^A \fd{\xi}^B  T_A T_B = \ha \fd{\xi}^A \fd{\xi}^B \lb T_A , T_B \rb = \ha \fd{\xi}^A \fd{\xi}^B C_{AB}{}^C T_C = \ha \lb \f{\Th}, \f{\Th} \rb
$$}
\beq
\ff{F} = \f{d} \, \f{\Th} + \f{\Th} \, \f{\Th} = \f{d} \, ( g^- \f{d} g) + g^- (\f{d} g) \, g^- \f{d} g = 0 \label{C2S}
\eeq
implying that Lie groups are, in this sense, perfectly symmetric.

The tangent space at every Lie group manifold point is spanned by the generator vector fields, $\ve{\xi}_A(x)$, at that point. Since, algebraically, the Lie derivative between these vector fields is the same as the commutator bracket between corresponding matrix Lie algebra generators, it is useful to consider the Lie algebra as the \emph{model tangent space} (also known as the ``internal space'' or ``fake tangent space''\footnote{The name ``fake tangent bundle'' was introduced casually by John Baez in \texttt{\href{http://math.ucr.edu/home/baez/week176.html}{TWF176}}, and used in~\cite{Wise2}.}), with the correspondence between tangent vectors at any point and vectors in the model tangent space determined by the Maurer-Cartan form, $\ve{\xi}_A \f{\Th} = T_A$. One often thinks of generator vector fields as Lie algebra generators, $\ve{\xi}_A \sim T_A$, but Lie algebra generators should not always be thought of as identical to generator vector fields. Rather, the Lie algebra, as the model tangent space, can be thought of as the ``rest frame,'' with the map of tangent vectors into this rest frame specified by the Maurer-Cartan form, thereby describing the local geometry of the Lie group manifold.

We may use the Lie bracket to build a natural bilinear form for the Lie algebra, the \emph{Killing form},
$$
(A,B) = \Tr_C [ [A,[B,C]]
$$
This produces a natural metric (also referred to as the Killing form) for the Lie algebra (the model tangent space) from the structure constants,
\beq
n_{AB} = \lp T_A, T_B \rp = C_{AC}{}^D C_{BD}{}^C \label{kf}
\eeq
which, for semisimple Lie algebras and a nice choice of generators, can consist of $+1$ or $-1$ entries on the diagonal. In practice, when working with a specific matrix group representation and coordinatization, for a semisimple Lie group, the Maurer-Cartan form components can be computed via
$$
\xi_i{}^A = \lp n^{AB} T_B, g^- \pa_i g(x) \rp
$$
using  (\ref{MC}) and the inverse of the Killing form, $n^{AB}$.

When the Killing form is nondegenerate, these Maurer-Cartan form components can be used to provide a pseudo-Riemannian metric over the Lie group manifold,
\beq
g_{ij} = \xi_i{}^A \, n_{AB} \, \xi_j{}^B \label{metric}
\eeq
With respect to this metric, the generator vector fields, $\ve{\xi}_A$, are both a vielbein and a set of $n$ Killing vector fields, corresponding to symmetries of the Lie group manifold. A \emph{Killing vector field}, $\ve{\xi}$, satisfies a version of Killing's equation,
$$
{\cal L}_{\ve{\xi}} \, \fd{\xi}^B = B_C{}^B \fd{\xi}^C 
$$
in which $B_C{}^B$, possibly position dependent, must be a rotation, $B_C{}^B = - B^B{}_C$, of the frame, $\fd{\xi}^B$, so that the metric is invariant under the flow of $\ve{\xi}$. Using the definition of the Lie derivative and the vanishing curvature of the Maurer-Cartan form (\ref{C2S}), Killing's equation for a generator vector fields is
$$
{\cal L}_{\ve{\xi}_A} \, \fd{\xi}^B = \ve{\xi}_A (\f{d}  \fd{\xi}^B) + \f{d} ( \ve{\xi}_A \fd{\xi}^B )
=  \ve{\xi}_A ( - \ha \fd{\xi}^D \fd{\xi}^C C_{DC}{}^B ) = -C_{AC}{}^B \fd{\xi}^C
$$
with $C_{AC}{}^B=-C_{A}{}^B{}_C$.

Another $n$ Killing vectors, $\ve{\chi}_A$, correspond to the left (rather than right) action of the Lie algebra generators,
\beq
T_A \, g = {\cal L}_{\ve{\chi}_A} g = \ve{\chi}_A \, \f{d} g = \chi_A{}^i \, \pa_i g(x) \label{chidef}
\eeq
and relate to the other generator vector fields by a rotation, $\ve{\chi}_A = L_A{}^B \ve{\xi}_B $, with
\beq
L_A{}^B (x) = \ve{\chi}_A \, \fd{\xi}^B = \ve{\chi}_A \, (T^B , g^- \f{d} g) = (T^B , g^- T_A g) \label{Ldef}
\eeq
These vector fields are also Killing with respect to the Maurer-Cartan form components,
$$
{\cal L}_{\ve{\chi}_A} \fd{\xi}^B = \ve{\chi}_A (\f{d}  \fd{\xi}^B) + \f{d} ( \ve{\chi}_A \fd{\xi}^B )
= \fd{\xi}^C L_A{}^D C_{CD}{}^B + \f{d} L_A{}^B
= 0
$$
and their mutual Lie derivatives produce the ``wrong'' sign, $\lb \ve{\chi}_A,\ve{\chi}_B \rb_{\cal L} = - C_{AB}{}^C \ve{\chi}_C$.

The Maurer-Cartan form may be used to define the \emph{Haar measure} for integration of functions over the Lie group manifold,
\beq
\int_G \nf{d^G \! x} \, f(x) = \int_G \fd{\xi}^1 \! \dots \fd{\xi}^n  f(x) = \int_G \nf{d^n \! x} \left| \xi \right| f(x) \label{Haar}
\eeq
producing, for example, the Lie group volume, $V=\int_G \nf{d^G \! x}$. The Maurer-Cartan form and Killing form also allow us to define a \emph{Hodge duality operator}, $\accentset{\Th}{\star}$, for forms over $G$.

There is a fundamental structure distinguishing a Lie group from an arbitrary manifold: Points of a Lie group correspond to manifold autodiffeomorphisms, via the left and right group product. Points near the identity correspond to flow diffeomorphisms, described by vector fields over the manifold, represented by matrix Lie algebraic generators. The geometric relationship between these flow vector fields corresponds to their mutual Lie derivative and to the commutator bracket between corresponding generators. The Maurer-Cartan form, a natural construct arising from these group product diffeomorphisms, describes how tangent vectors anywhere on the manifold correspond to flow vector fields and Lie algebra elements.

From a geometric perspective, every point on a Lie group manifold looks the same, with flow vector fields in all directions, corresponding to generator vectors in the model tangent space, with Lie derivatives between flow vector fields having the same structure at each point. For a \emph{Lie group manifold}, ${\cal G}$, unlike for a Lie group, $G$, there is nothing special about the identity point. What we have been calling a ``Lie group manifold,'' which is also called a ``torsor,'' ``G-torsor,'' or ``principal homogeneous space,'' is a Lie group that has forgotten where its identity point is. An $n$-dimensional Lie group manifold, with a specific topology and its natural Maurer-Cartan form, has the same local geometry and global topology as the corresponding Lie group. Physicists often fail to draw a distinction between a Lie group and the corresponding Lie group manifold, and this distinction can usually be surmised from context. For example, a Lie group manifold serves as the typical fiber when constructing a principal fiber bundle. 

\section{Fiber bundles with connections}

All fundamental fields of physics---the scalar Higgs field, electroweak and strong gauge fields, gravitational spin connection and frame, all fermions, and ghosts---are conventionally described as connections on or sections of a fiber bundle. Geometrically, a fiber bundle consists of a base manifold, $M$, usually considered as our spacetime, with a copy of a typical fiber, $F$, over each point. The base and fibers combined constitute the \emph{total space} (or ``entire space'') manifold, $E$, with a defining map, $\pi : E \to M$, projecting to the base. The defining map determines a fiber submanifold of $E$ over each point of $M$. We can cover the base with patches, $U_a$, such that the patches of the total space over these are a direct product with the typical fiber space, $E_{U_a} \simeq U_a \times F$, with coordinates, $(x,y)$, of points in $E_{U_a}$ corresponding to $x$ in $U_a$ of $M$ and $y$ in $F$. Fibers over overlapping patches of the base are stitched together using some action of the \emph{structure group}, $G$, on the fiber. Each different kind of field in physics corresponds to a different typical fiber, each fiber a representation space of a Lie group. A specific field configuration, $\phi_a(x)$, valued in a fiber over a patch of the base corresponds to a \emph{local section}, $\si_a :U_a \to E_{U_a}$, of the fiber bundle, satisfying $\pi(\si_a(x)) = x$, so that, in coordinates, $\si_a (x) = (x,\phi_a(x))$. For most cases in physics, with $M$ a typical four-dimensional spacetime, the total space is trivial, $E \simeq M \times F$, and admits global sections, $\si : M \to E$, mapping $M$ to submanifolds of $E$ specified by $\si (x) = (x,\phi(x))$. Sections of a trivial bundle are equivalent to continuous functions, $\ph(x)$, from $M$ into $F$.

With coordinate dependent elements of the structure group, $G$, called \emph{transition functions}, used to relate fibers over overlapping patches of $M$, it is possible to use the same transition functions on \emph{associated fibers} that are also representation spaces of $G$. The most important case to consider is the associated \emph{principal fiber bundle}, $P$, over $M$, for which the associated fiber is the Lie group manifold, ${\cal G}$, corresponding to the structure group, with transition functions acting via the left action.

To describe how a field, $\ph(x) \in F$, valued in some representation space changes from point to nearby point in spacetime, nearby fibers are related by a \emph{connection},
$$
\f{A} = \f{dx}^i A_i^B(x) \, T_B
$$
a $1$-form field over $M$ valued in the Lie algebra of the structure group, $Lie(G)$. This is used to construct a \emph{covariant derivative},
$$
\f{D} \ph = \f{d} \ph + \f{A} \ph
$$
For any field configuration a \emph{gauge-equivalent} field, $\ph'(x) = g^-(x) \, \ph(x)$, (and a gauge-equivalent section of the fiber bundle) is given by transforming by any chosen $G$-valued function of position over $M$. (Here, inverse group elements, $g^-(x)$, are used to conform with subsequent conventions.) Under this \emph{gauge transformation}, $\ph(x) \mapsto \ph'(x) = g^-(x) \, \ph(x)$, the covariant derivative transforms to $\f{D}' \ph' = g^- \f{D} \ph$, so the connection must transform to
$$
\f{A}' = g^- \f{A} g + g^- \f{d} g
$$
Any connection defined over $M$ corresponds to a specific field over the entire associated principal fiber bundle, $P$. For any section, $\si'(x) = (x,y(x))$, of the principal fiber bundle corresponding to a chosen gauge transformation, $g^-(x) = g^-(y(x))$, the connection, $\f{A}'$, over $M$, is the pullback, $\f{A}' = \si'^* \f{\cal A}$, of an \emph{Ehresmann connection form}, $\f{\cal A}$---a $Lie(G)$-valued $1$-form field over $P$ which may be expressed using local coordinates as
\beq
\f{\cal A}(x,y) = g^-(y) \f{A}(x) g(y) + g^-(y) \f{d^y} g(y) \label{ecf}
\eeq
In physics, quantities dependent on the connection or covariant derivative are often \emph{gauge invariant}---independent of the choice of section of $P$. Using (\ref{Ldef}) and identifying the Maurer-Cartan form along the principal fibers, the Ehresmann connection form (\ref{ecf}) may also be written as
$$
\f{\cal A}(x,y) = \f{A}^B(x) L_B{}^C(y) T_C  + \f{\Th}(y)
$$
By relating the Lie algebra generators, $T_A$, to the corresponding vector fields, $\ve{\xi}_A(y)$, along the principal fibers using the Maurer-Cartan form, $\f{\cal A}(x,y)=\f{\ve{\cal A}}(x,y) \f{\Th}(y)$, we may define the \emph{Ehresmann connection} locally as a vector-valued $1$-form over $P$,
$$
\f{\ve{\cal A}}(x,y) = \f{A}^B(x) L_B{}^C(y) \ve{\xi}_C(y)  + \fd{\xi}^C(y) \ve{\xi}_C(y) = \f{A}^B(x) \ve{\chi}_B(y)  + \f{\ve{\Th}}(y)
$$
in which $\ve{\chi}_B$ are the left-action vector fields and $\f{\ve{\Th}}(y) = \fd{\xi}^C(y) \ve{\xi}_C(y) = \fd{dy^p} \ve{\pa}_p$ is the identity projection along principal fibers. As a natural, globally defined vector projection, $\f{\ve{\cal A}} \, \f{\ve{\cal A}} = \f{\ve{\cal A}}$, the Ehresmann connection maps vectors on the total space, $P$, to their projections along the tangent space to the principal fibers.

In more detail, the Ehresmann connection splits the tangent bundle of the principal fiber bundle into \emph{vertical} and \emph{horizontal} distributions, satisfying $\ve{\De}_V \f{\ve{\cal A}} = \ve{\De}_V$ and $\ve{\De}_H \f{\ve{\cal A}} = 0$. The vertical distribution, spanned at each point of the principal fiber bundle by coordinate basis vectors along the fibers, $\ve{\pa}_y \in \ve{\De}_V$, integrates to principal fiber submanifolds of the total space. The horizontal distribution, spanned at each point of the principal fiber bundle by four horizontal vectors at each point,
$$
\ve{h}_i = \ve{\pa}_i - A_i{}^B \ve{\chi}_B \in \ve{\De}_H
$$
integrates to submanifolds of the total space with $\ve{h}_i$ as tangent vectors if and only if the curvature of the connection vanishes. In this way, the Ehresmann connection, a natural vector-valued $1$-form over the total space, describes the geometry of the principal fiber bundle. The Ehresmann connection and corresponding decomposition of the tangent bundle into horizontal and vertical distributions are the fundamental structures of \emph{Ehresmannian geometry}.

A generalization of the Lie bracket to vector-valued forms, called the ``Fr\"{o}licher-Nijenhuis bracket'' (or ``FuN bracket''), allows the definition of the \emph{FuN curvature} of the Ehresmann connection over the total space,
\begin{eqnarray*}
\ff{\ve{\cal F}}(x,y) &=& - \ha \lb \f{\ve{\cal A}}, \f{\ve{\cal A}} \rb_{\cal L}
= - \f{\ve{\cal A}} \lp \f{\pa} \f{\ve{\cal A}} \rp + \f{\pa} \lp \f{\ve{\cal A}} \f{\ve{\cal A}} \rp
= \lp 1 - \f{\ve{\cal A}} \rp \lp \lp \f{\pa}^x + \f{\pa}^y \rp \f{\ve{\cal A}} \rp \\
&=& \lp 1 - \f{A}^B(x) \ve{\chi}_B(y) - \fd{dy^p} \ve{\pa}_p \rp 
\lp \lp \f{\pa}^x \f{A}^C \rp \ve{\chi}_C(y) - \f{A}^C \f{\pa}^y \ve{\chi}_C \rp \\
&=& \lp \f{\pa}^x \f{A}^C \rp \ve{\chi}_C - \f{A}^B \f{A}^C \ve{\chi}_B \f{\pa}^y \ve{\chi}_C \\
&=& \lp \f{\pa}^x \f{A}^D + \ha \f{A}^B \f{A}^C C_{BC}{}^D \rp \ve{\chi}_D(y) \\
&=& \ff{F}{}^D (x) \ve{\chi}_D(y)
\end{eqnarray*}
matching the curvature of the Ehresmann connection form,
$$
\ff{\cal F} = \f{d} \f{\cal A} + \f{\cal A} \f{\cal A} = g^-(y) \ff{F}(x) g(y)
$$
in which
$$
\ff{F}(x) =  \ff{F}{}^D T_D = \f{d} \f{A} + \f{A} \f{A} = \f{D} \f{A}
$$
is the curvature of the connection over $M$.

Although we motivated the existence of a principal fiber bundle connection by the necessity of describing a covariant derivative for an associated fiber bundle, this motivation is not essential. A principal fiber bundle with connection, by itself, is an interesting geometric object, with curvature specified by the covariant derivative of the connection. The choice of which kind of connection to consider more fundamental---the standard connection, $\f{A}$, over $M$, the Ehresmann connection form, $\f{\cal A}$, over $P$, or the Ehresmann connection, $\f{\ve{\cal A}}$---is largely a matter of taste. Physicists typically use $\f{A}$ in calculations, are vaguely aware of $\f{\cal A}$, and may have been in the same room with $\f{\ve{\cal A}}$. Mathematicians, on the other hand, typically favor $\f{\ve{\cal A}}$ or its more abstract description as a splitting of the tangent space into horizontal and vertical distributions. The main motivation for favoring $\f{\ve{\cal A}}$ is that it is completely natural, consisting of a map from vector fields to vector fields, and defined globally over $P$.

Even though the Maurer-Cartan form (\ref{MC}) is usually thought of as the frame over a Lie group manifold, it can also be considered as the connection of a special principal fiber bundle. In a trivial way, the Maurer-Cartan form is the Ehresmann connection form of a $G$-bundle with a zero-dimensional base, $\f{\cal A}(y)=\f{\Th}(y)=\fd{\xi}^B(y) T_B$. But the corresponding Ehresmann connection is just the identity map, $\ve{\f{\Th}}(y) = \fd{dy^p} \ve{\pa_p} = \ve{\f{1}}$, since the entire tangent space of that bundle is vertical. To fully describe the local geometry of a Lie group manifold naturally, with an Ehresmann-Maurer-Cartan connection, we must consider a more interesting principal bundle.

We take both the $n$-dimensional base manifold, $M$, and the typical fiber, $F$, to be copies of the Lie group manifold, forming the $2n$-dimensional \emph{principal total space}, $P_{\cal G}  = {\cal G} \times {\cal G}$. The \emph{Maurer-Cartan connection} is then a specific Lie algebra-valued connection over the $n$ dimensional base,
\beq
\f{A}(x) = \f{\Th}(x) =  \fd{\xi}^B(x) \, T_B \label{MCc}
\eeq
with the corresponding \emph{Ehresmann-Maurer-Cartan connection form} and \emph{Ehresmann-Maurer-Cartan connection} defined over $P_{\cal G}$,
\begin{eqnarray}
\f{\it \Th}(x,y) &=& g^-(y) \f{\Th}(x) g(y) + \f{\Th}(y)
= \fd{\xi}^B(x) L_B{}^C(y) T_C  +  \fd{\xi}^C(x) T_C = \f{\ve{\it \Th}}(x,y) \, \f{\Th}(y) \\[.5em]
\f{\ve{\it \Th}}(x,y) &=& \fd{\xi}^B(x) L_B{}^C(y) \ve{\xi}_C(y) + \fd{\xi}^C(y) \ve{\xi}_C(y)
= \fd{\xi}^B(x) \ve{\chi}_B(y)  + \f{\ve{\Th}}(y) \label{EMCc}
\end{eqnarray}
In this completely natural Ehresmann-Maurer-Cartan connection, $\f{\ve{\it \Th}}$, we see the local geometry of a Lie group described via a map from right-action generator vector fields on the base ${\cal G}$ to the corresponding left-action generator vector fields on the ${\cal G}$ fibers. The principal total space, $P_{\cal G}={\cal G} \times {\cal G}$, is necessary for naturally describing the local geometry of a Lie group via the Ehresmann-Maurer-Cartan connection, $\f{\ve{\it \Th}}$, corresponding to the Maurer-Cartan connection, $\f{\Th}$, over the Lie group manifold base, ${\cal G}$. This construction makes precise and natural the previously introduced notion of the Maurer-Cartan form as a map of tangent vectors into the model tangent space, now understood as the tangent space of the typical fiber of the principal total space. More succinctly, the Maurer-Cartan form (\ref{MC}) is a specific map of tangent vectors on the base ${\cal G}$ to tangent vectors on a fiber ${\cal G}$, identified as Lie algebra elements. Although this description is completely natural, working with the Ehresmann-Maurer-Cartan connection, $\f{\ve{\it \Th}}$, over $P_{\cal G}$ is cumbersome, and we will usually prefer to work with and think about the equivalent Maurer-Cartan connection, $\f{\Th}$, over ${\cal G}$, valued in the Lie algebra---the model tangent space. This preference is motivated by the similar construction and description of pseudo-Riemannian geometry and gravity using a frame.

\section{Gravity}

The fiber bundle structure---Ehresmannian geometry---is well suited for describing curving spacetime and gravitational physics. Physically, we move around in a four-dimensional base manifold, $M$, in which, at each point, $x$, we can imagine a freely falling clock and set of three orthogonal rulers. Each clock moves along a path in $M$ with unit tangent vector, $\ve{e}_0(x)$, corresponding to durations of time. The three rulers are tangent to paths with unit tangent vectors, $\ve{e}_\pi(x)$, corresponding to directions and magnitudes of distance. These four orthonormal vector fields, $\ve{e}_\mu(x)$, constitute a \emph{tetrad} for spacetime, implying the existence of a pseudo-Riemannian metric, $g_{i j}(x)$, such that, locally,
$$
(\ve{e}_\mu, \ve{e}_\nu) = e_\mu{}^i \, g_{ij} \, e_\nu{}^j = \eta_{\mu \nu}
$$
in which $\eta$ is the Minkowski metric, presumed to have $\eta_{00}=+1$. The matrix of tetrad components, $e_\mu{}^i$, is invertible, implying the existence of four $1$-forms, the \emph{cotetrad}, $\f{e}^\mu$, satisfying
$$
\ve{e}_\mu \f{e}^\nu = e_\mu{}^i \ve{\pa}_i \f{dx}^j e_j{}^\nu = e_\mu{}^i \de_i^j e_j{}^\nu = e_\mu{}^i e_i{}^\nu = \de^\nu_\mu
$$
The cotetrad resolves tangent vectors into components in the freely falling reference frame of the standard clock and rulers, $\ve{v} \f{e}^\mu = v^\mu$. With a tetrad defining frames of reference, we may associate the temporal and spatial tetrad vectors with algebraic elements, $\ve{e}_\mu \sim \ga_\mu$, spanning a model tangent space and satisfying the same orthonormality condition for some bilinear form, $(\ga_\mu,\ga_\nu) = \eta_{\mu \nu}$. The Clifford algebra $Cl(1,3)$ is ideally suited to this purpose, with the four $\ga_\mu$ identified as Clifford basis vectors satisfying
$$
(\ga_\mu,\ga_\nu) = \ga_\mu \cdot \ga_\nu = \ha ( \ga_\mu \ga_\nu + \ga_\nu \ga_\mu )  = \eta_{\mu \nu}
$$
With these algebraic elements we can define the spacetime \emph{frame} to be a Clifford vector-valued $1$-form,
$$
\f{e} = \f{e}^\mu(x) \, \ga_\mu
$$
mapping tangent vectors on $M$ into the model tangent space spanned by the Clifford basis vectors,
$$
\ve{v} \f{e} = v^i \ve{\pa}_i \f{dx}^j e_j{}^\mu \ga_\mu = v^\mu \ga_\mu = v \,\,\, \in \, Cl^1(1,3)
$$
The Clifford algebra basis elements, such as $\ga_\mu$, may be represented by matrices, such as the Dirac matrices, with the Clifford product isomorphic to the matrix product.

Under General Relativity there is no preferred rest frame at each point---we are free to choose a different but gauge-equivalent frame, $\f{e}' = g^- \f{e} g$, related to the original by some rotation and boost. Interpreted as a section of a fiber bundle, the frame has $Cl^1(1,3)$ as typical fiber and $Spin(1,3)$ as structure group. The \emph{spin connection} for this bundle can be written as a Clifford bivector-valued $1$-form,
\beq
\f{\om} = \f{dx}^i \ha \om_i{}^{\mu \nu}(x) \, \ga_{\mu \nu} \label{spinc}
\eeq
in which six independent $spin(1,3)$ basis generators\footnote{We will typically write Lie algebras using lower case, $spin(p,q) = Lie(Spin(p,q))$.} are Clifford bivectors,
$$
\ga_{\mu \nu} = \ga_\mu \times \ga_\nu = \ha ( \ga_\mu \ga_\nu - \ga_\nu \ga_\mu ) \,\,\, \in \, Cl^2(1,3) \sim spin(1,3) = Lie(Spin(1,3))
$$
Using this connection the covariant derivative of the frame is the \emph{torsion},
\beq
\ff{T} = \f{D} \f{e} = \f{d} \f{e} + \f{\om} \times \f{e} = \f{dx}^i \f{dx}^j ( \pa_i e_j{}^\mu + \om_i{}^\mu{}_\nu e_j{}^\nu ) \, \ga_\mu \label{T}
\eeq
in which we have used the antisymmetric Clifford product,
\beq
\ga_{\mu \nu} \times \ga_\rh = \ha ( \ga_{\mu \nu} \ga_\rh - \ga_\rh \ga_{\mu \nu} ) = \et_{\nu \rh} \ga_\mu - \et_{\mu \rh} \ga_{\nu} \label{biv}
\eeq
The \emph{Riemann curvature}, as a Clifford bivector-valued (or $spin(1,3)$-valued) $2$-form, is
\beq
\ff{R} = \f{d} \f{\om} + \ha \f{\om} \f{\om}
= \f{dx}^i \f{dx}^j \ha ( \pa_i \, \om_j{}^{\mu \nu} + \om_i{}^\mu{}_\rh \, \om_j{}^{\rh \nu}) \, \ga_{\mu \nu} \label{R}
\eeq
in which we have used the antisymmetric Clifford product again, here equivalent to the $spin(1,3)$ bracket,
$$
\ha [ \ga_{\mu \nu}, \ga_{\rh \si} ] = \ga_{\mu \nu} \times \ga_{\rh \si}
= \et_{\nu \rh} \ga_{\mu \si} - \et_{\nu \si} \ga_{\mu \rh} - \et_{\mu \rh} \ga_{\nu \si} + \et_{\mu \si} \ga_{\nu \rh}
$$
Complementing the antisymmetric product, the symmetric Clifford product of bivectors produces a scalar and quadvector, with the scalar giving the $spin(1,3)$ Killing form,
\begin{eqnarray*}
\ga_{\mu \nu} \cdot \ga_{\rh \si} &=& (\ga_{\mu \nu}, \ga_{\rh \si}) + \ga_{\mu \nu \rh \si} \\
(\ga_{\mu \nu}, \ga_{\rh \si}) &=& \et_{\mu \si} \et_{\nu \rh} - \et_{\mu \rh} \et_{\nu \si}
\end{eqnarray*}
In General Relativity we also encounter the \emph{Ricci curvature}, $\f{R} = \ve{e} \times \ff{R}$, a vector-valued $1$-form, and the \emph{curvature scalar}, $R = \ve{e} \cdot \f{R}$, in which we use the \emph{coframe}, $\ve{e} = \ga^\mu \, \ve{e}_\mu (x)$.

In the $Cl(1,3)$ Clifford algebra, the single quadvector basis element is the \emph{pseudoscalar},
$$
\ga = \ga_0 \ga_1 \ga_2 \ga_3
$$
which gives, for example, $\ga_{\mu \nu \rh \si} = \ep_{\mu \nu \rh \si} \ga$, using the permutation symbol. Multiplication (on the right) by the pseudoscalar (or, technically, its inverse, $\ga^- = - \ga$) provides a natural \emph{duality} operation within a Clifford algebra. In this way, the Clifford dual of any grade $r$ element is a grade $(4-r)$ element,
$$
A^r \ga^- = \fr{1}{r!} A^{\al \dots \be} \ga_{\al \dots \be} \ga^- = \fr{1}{r! \lp 4 - r \rp !} A^{\al \dots \be} \lp \ga_{\al \dots \be \ga \dots \de} \ga^- \rp \ga^{\ga \dots \de} = \fr{1}{r! \lp 4 - r \rp !} A^{\al \dots \be} \ep_{\al \dots \be \ga \dots \de} \ga^{\ga \dots \de}
$$ 
in which the Clifford indices have been raised using the Minkowski metric, $\ga^\mu = \eta^{\mu \nu} \ga_\nu$. In general, the Clifford dual of a $r$-vector is a $(n-r)$-vector in a dimension $n$ Clifford algebra. Applied to spacetime bivectors, Clifford duality provides the duality operation for the $spin(1,3)$ Lie algebra, such as $\ga_{01} \ga^- = \ga_{23}$. Also, we can use the spacetime coframe, $\ve{e}$, to map any differential $p$-form, $\nf{a} = \fr{1}{p!} a_{i \dots j} \f{dx}^i \dots \f{dx}^j$, to a Clifford $p$-vector, compute its Clifford dual, then use the spacetime frame to convert back, obtaining the \emph{Hodge dual} $(4\!-\!p)$-form,
$$
\nf{\star a} = \fr{1}{p! \lp 4-p \rp!} a_{\al \dots \be} \ep^{\al \dots \be \ga \dots \de} \f{e}_\ga \dots \f{e}_\de
= \fr{\sqrt{\ll g \rl}}{p! \lp 4-p \rp!} a_{i \dots j} g^{im} \dots g^{jn} \ep_{m \dots n k \dots l} \f{dx}^k \dots \f{dx}^l = \star \nf{a}
$$
The \emph{Hodge star} operator, $\star$, acts on forms to convert them to their Hodge dual. This operator requires the existence of a spacetime frame, or at least a metric, and generalizes to work in higher dimensional spacetimes. When working in various manifolds it is important to be mindful of which metric is being used to build which Hodge operator, especially since the spacetime Hodge operator is needed for describing Yang-Mills dynamics. For a $2$-form, $\ff{F} = \ha \f{e^\mu} \f{e^\nu} F_{\mu  \nu}$, in $4$-dimensional spacetime, its Hodge dual is the $2$-form, $\ff{\star F} = \fr{1}{4} F_{\mu \nu} \ep^{\mu \nu \rh \si} \f{e}_\rh \f{e}_\si$. For the spacetime area $2$-form, $\f{e}\f{e}$, its Hodge dual is the same as its Clifford dual,
\begin{equation}
\star \f{e} \f{e} = \ha \ga_\mu \ga_\nu \ep^{\mu \nu \rh \si} \f{e}_\rh \f{e}_\si = \ha \f{e}^\mu \f{e}^\nu \ep_{\mu \nu \rh \si} \ga^\rh \ga^\si = \f{e}^\mu \f{e}^\nu \ga_\mu \ga_\nu \ga^- = \f{e} \f{e} \ga^- \label{eeg}
\end{equation}
a cute duality identity, useful for describing gravity. 

\section{De Sitter cosmology}

A specific example of a realistic spacetime cosmology, describing a universe approximating ours, with exponentially expanding flat spatial sections, is \emph{de Sitter spacetime}, with frame
\beq
\f{e}^{\mbox{\tiny S}} = \f{dx}^i e^{\mbox{\tiny S}}_i{}^\mu \ga_\mu =  \f{dt} \ga_0 + \f{d s^\pi} e^{\al t} \ga_\pi \label{dSe}
\eeq
Here, $\al$ is the \emph{expansion parameter}, and the unit-carrying physics coordinates are the time measured by a standard clock, $t = x^0 T$, with $T$ being some temporal unit such as seconds, and spatial lengths, $s^\pi = x^\pi (L/c)$, in units of time, with $L$ a spatial unit, such as meters, and $c$ the speed of light. Mathematically, manifold coordinate labels, $x^i$, have no units, and it is the frame components, $e^{\mbox{\tiny S}}_i{}^\mu$, or scaled coordinates such as $t$, that have temporal units. Inverting the frame, the coframe is
$$
\ve{e}_{\mbox{\tiny S}} = \ga^0 \ve{\pa}_t + \ga^\pi e^{-\al t} \ve{\pa}_\pi
$$
The torsionless ($\ff{T}=0$ in (\ref{T})) spin connection compatible with the de Sitter spacetime frame is
\beq
\f{\om}^{\mbox{\tiny S}} = - \f{ds}^\pi \al e^{\al t} \ga_{0 \pi} \label{dSw}
\eeq
with resulting Riemann curvature
\beq
\ff{R}^{\mbox{\tiny S}} = - \f{dt} \, \f{ds}^\pi \al^2 e^{\al t} \ga_{0 \pi} - \f{ds}^\rh \f{ds}^\pi \ha \al^2 e^{2 \al t} \ga_{\rh \pi} = - \fr{\al^2}{2} \f{e}^{\mbox{\tiny S}} \f{e}^{\mbox{\tiny S}} \label{dSR}
\eeq
Ricci curvature $\f{R}^{\mbox{\tiny S}} = -3 \al^2 \f{e}^{\mbox{\tiny S}}$, and scalar curvature $R^{\mbox{\tiny S}} = - 12 \al^2$. Physically, a de Sitter universe corresponds to empty space with expansion driven by a \emph{cosmological constant}, $\La = 3 \al^2$. As a spacetime of constant curvature, this vacuum solution satisfies Einstein's equations of General Relativity, and it approximates our physical universe, with large scale observations currently indicating $\La \simeq 1 \times 10^{-35} \fr{1}{s^2}$ and an expansion parameter of $\al \simeq 2 \times 10^{-18} \fr{1}{s}$. Of course, the existence of matter disturbs spacetime away from this de Sitter vacuum state.

\section{MacDowell-Mansouri gravity}

Most modern descriptions of gravity begin with the spin connection, $\f{\om}$, of a $Spin(1,3)$ principal bundle as the primary dynamical field, accompanied by the frame, $\f{e}$, of an associated fiber bundle, over a four-dimensional base. An interesting structural unification, first introduced by MacDowell and Mansouri, is achieved by combining spin connection and frame fibers as parts of a $Spin(1,4)$ principal bundle. This unification works because the antisymmetric Clifford product of $Cl^2(1,3)$ bivectors and $Cl^1(1,3)$ vectors (\ref{biv}) is isomorphic to the Lie bracket between corresponding $spin(1,4)$ generators,
$$
\ha [ \ga_{\mu \nu}, \ga_{\rh 4} ] = \ga_{\mu \nu} \times \ga_{\rh 4}
= \et_{\nu \rh} \ga_{\mu 4} - \et_{\mu \rh} \ga_{\nu 4}
= ( \ga_{\mu \nu} \times \ga_\rh ) \, \ga_4
$$
Using this algebraic identification of different parts of
$$
spin(1,4) \sim Cl^2(1,4) = Cl^2(1,3) \oplus Cl^1(1,3)
$$
the connection for the $Spin(1,4)$ bundle can be split into different parts,
\beq
\f{H} = \ha \f{\om} + \f{E} \label{bc}
\eeq
in which $\f{\om}$ is the $spin(1,3)$-valued part, identified with the spin connection (\ref{spinc}), and $\f{E}$ is valued in its complement,
$$
\f{E} = \f{E}^{\mu 4} \ga_{\mu 4} = \f{e}^\mu \fr{\al}{2} \ga_\mu \ga_4 = \f{e} \, \ph_{\mbox{\tiny S}}  
$$
with $\f{e}=\f{e}^\mu \ga_\mu$ identified as the frame, valued in a four-dimensional $Cl^1(1,3)$ subspace of $Cl^1(1,4)$, and
\beq
\ph_{\mbox{\tiny S}} = \ph^4_0 \ga_4 = \fr{\al}{2} \ga_4 \label{vev}
\eeq
a constant $Cl^1(1,4)$ \emph{Higgs vacuum expectation value} vector that commutes with the chosen $spin(1,3)$. Since the frame carries units of time, $[\f{e}]=T$, and connections are unitless, the Higgs and expansion parameter, $\al$, have units of inverse time, $[\ph_{\mbox{\tiny S}}]=[\al]=1/T= [c^2 / \hbar] M$, or mass in natural units. With this decomposition of $\f{H}$ into different parts, the $Spin(1,4)$ bundle curvature is
\beq
\ff{F} = \f{d} \f{H} + \f{H} \f{H} = (\ha \ff{R} + \fr{\al^2}{4} \f{e} \f{e}) + \ff{T} \ph_{\mbox{\tiny S}} \label{SF}
\eeq
in which $\ff{R}$ is the Riemann curvature (\ref{R}) and $\ff{T}$ is the torsion (\ref{T}). As shown by MacDowell and Mansouri and others, this curvature can be used to construct a realistic action for gravity.

If we consider de Sitter spacetime in the context of MacDowell and Mansouri's formulation, we find an interesting result. Because de Sitter spacetime (\ref{dSR}) is torsionless and has Riemann curvature $\ff{R}^{\mbox{\tiny S}} = - \fr{\al^2}{2} \f{e}^{\mbox{\tiny S}} \f{e}^{\mbox{\tiny S}}$, the corresponding $Spin(1,4)$ bundle curvature (\ref{SF}) vanishes, $\ff{F}^{\mbox{\tiny S}}=0$. The $Spin(1,4)$ bundle with connection corresponding to de Sitter spacetime is, in this sense, flat. This $spin(1,4)$-valued \emph{de Sitter connection}, $\f{H}^{\mbox{\tiny S}}$, is the combination (\ref{bc}) of the spin connection (\ref{dSw}) and frame (\ref{dSe}) of de Sitter spacetime,
\begin{eqnarray}
\f{H}^{\mbox{\tiny S}} &=&  \ha \f{\om}^{\mbox{\tiny S}} + \f{e}^{\mbox{\tiny S}} \ph_{\mbox{\tiny S}} \notag \\
&=& - \f{ds}^\pi \ha \al e^{\al t} \ga_{0 \pi} + \f{dt} \ha \al \ga_{04} + \f{d s^\pi} \ha \al e^{\al t} \ga_{\pi 4} \notag \\
&=& \f{dt} \, \al N_0 + \f{d s^\pi} \al e^{\al t} N_\pi \label{dsc}
\end{eqnarray}
in which
\beq
N_0 = \ha \ga_{04} \;\;\;\;\;\; \mathrm{and} \;\;\;\;\;\; N_\pi = \ha (\ga_{\pi 4} - \ga_{0 \pi}) = \ha \ga_\pi (\ga_4 + \ga_0) \label{tngen}
\eeq
are the temporal and three null \emph{de Sitter generators}. Algebraically, these null generators are eigenvectors (root vectors) of the temporal generator, satisfying
\begin{align}
\lb N_0, N_\pi \rb &= - N_\pi & N_0 N_\pi &= - \ha N_\pi & \lp N_0, N_0 \rp &= N_0 N_0 = \fr{1}{4} \notag \\
\lb N_\pi, N_\rh \rb &= 0 & N_\pi N_\rh &= 0 & \lp N_\pi, N_\rh \rp &= 0 
\label{nident}
\end{align}
and will play an important role in what follows.

Although combining the spin connection and frame as parts of a larger connection works perfectly as an algebraic unification, this construction introduces an important mystery. If the fundamental structure of our universe is presumed to be a $Spin(1,4)$ principal bundle with connection over a four-dimensional base, then there is no natural reason why part of the $spin(1,4)$-valued connection should relate to the spacetime frame over the base. We arbitrarily split the connection by hand to obtain a part that we then treated as a spacetime frame, without that split or identification being justified. Even if we contrived some symmetry breaking scheme to accomplish the splitting of the connection, that still would not justify identifying part of the connection as the spacetime frame---a serious foundational inadequacy of the MacDowell-Mansouri approach to gravity. This ontological malady can be cured by considering a slightly different geometric framework, \emph{Cartan geometry}.

\section{Cartan geometry}

The idea behind Cartan geometry is to deform (or ``excite'') a Lie group while preserving a chosen subgroup, so the Lie group becomes a principal bundle with a connection and associated frame. We begin by factoring a Lie group, $G$, into a subgroup, $H$, and a \emph{coset space}, $G/H$, obtained by modding $G$ by $H$. The resulting factored geometry---the Lie group, $G \simeq G/H \times H$, decomposed into these parts---is called a \emph{Klein geometry}. More specifically, the ($dim(G) - dim(H)$ dimensional) coset space consisting of points, $x \in G/H$, is related to submanifolds of $G$ specified by \emph{coset representatives}, $r(x) \in G$, each in an equivalence class under right action by $H$,
$$
x \sim \lb r(x) \rb = \lb r(x) \, h(y) \rb =  r(x) \, H = \left\{ r(x) \, h(y) : \forall \, h(y) \in H \right\}
$$
A coset space manifold without the identity point specified is a \emph{homogeneous space}, also labeled $G/H$, on which $G$ acts continuously and transitively. The homogeneous space, $G/H$, may be considered the base of a principal bundle with $H$ as typical fiber and the original Lie group, $G$, as the total space, with defining map $\pi : g \mapsto [g]$. The chosen coset representative section, $r : G/H \rightarrow G$, serves as a reference section and local trivialization of this bundle. Gauge equivalent sections, $r'(x) = r(x) \, h(x)$, are related via gauge transformations by $H$-valued functions. The pullback of $G$'s Maurer-Cartan form, $\f{\it \Th}(x,y) = g^- \f{d} g$, on this section produces the \emph{Maurer-Cartan connection} for the Klein geometry,
$$
\f{\Th}(x) = r^* \f{\it \Th} = r^- \f{d} r = \f{A}^{\mbox{\tiny S}} + \f{E}^{\mbox{\tiny S}}  \,\,\, \in \, Lie(G) = Lie(H) \oplus Lie(G/H)
$$
The Maurer-Cartan connection consists of a specific H-connection, $\f{A}^{\mbox{\tiny S}}$, valued in the Lie algebra of the subgroup, $Lie(H)$, and a second part, the frame for the homogeneous space, $\f{E}^{\mbox{\tiny S}}$, valued in $Lie(H)$'s complement, $Lie(G/H)$. A Klein geometry is a decomposition of a Lie group and its Maurer-Cartan form as the total space of a principal bundle with a subgroup as typical fiber and a specific connection and frame over a homogenous base space.

With a Lie group factored into a Klein geometry, integration of functions over the Lie group can be separated as
\beq
\int_G \nf{d^G \! z} \, f(z) = \int_{G/H} \nf{d^{G/H} \! x} \, \int_H \nf{d^H \! y} \, f ( x,y) \label{sepi}
\eeq
in which $\nf{d^G \! z}$ is the Haar measure (\ref{Haar}) on $G$, $\nf{d^H \! y}$ is the Haar measure on $H$, $\nf{d^{G/H} \! x}$ is the measure on $G/H$ provided by the frame part of the Maurer-Cartan connection, $\f{E}^{\mbox{\tiny S}}$, multiplied by a constant scale factor, and $f(z) = f(x,y)$ is the integrand as a function of Lie group points, $z$, coordinatized by $x$ over $G/H$ and $y$ over $H$.

A \emph{Cartan geometry} is constructed by allowing a Klein geometry to \emph{deform}, or be excited, with the Maurer-Cartan connection allowed to vary arbitrarily, becoming a \emph{Cartan connection},
\beq
\f{C}(x) = \f{A} + \f{E} \label{Carcon}
\eeq
consisting of an H-connection, $\f{A}$, valued in $Lie(H)$, and frame, $\f{E}$, valued in $Lie(G/H)$, over a base space, $M$, \emph{modeled} on the homogeneous space, $G/H$. This Cartan geometry describes a deformation of the original Lie group, $G$, to $\tilde{G} \simeq M \times H$, in which subgroup fibers, $H$, maintain their geometry. The resulting structure is a principal H-bundle over a base, $M$, having the same dimensionality as $G/H$, along with an $H$-connection and frame over $M$. The reference section, $r:G/H \to G$, of the Klein geometry becomes a section, $\si:M \to \tilde{G}$, of the total space of the principal H-bundle, now seen as a deforming $\tilde{G}$. The H-connection and frame parts of the Cartan connection over M are the pullback, $\f{C}= \si^* \f{\cal C}$, of the Ehresmann connection form (\ref{ecf}) and \emph{Ehresmann frame form} parts of the \emph{Ehresmann-Cartan connection form} over $\tilde{G}$,
\beq
\f{\cal C}(x,y) = \f{\cal A}(x,y) + \f{\cal E}(x,y) = \lp h^-(y) \f{A}(x) h(y) + h^-(y) \f{d^y} h(y) \rp +  h^-(y) \f{E}(x) h(y)  \label{ECc}
\eeq
The curvature of the Cartan geometry is
\beq
\ff{\cal F}(x,y) = \f{d} \f{\cal C} + \f{\cal C} \f{\cal C} = h^-(y) \ff{F}(x) h(y) \label{Fcov}
\eeq
in which
\beq
\ff{F}(x) = \f{d} \f{C} + \f{C} \f{C}
\eeq
is the curvature of the Cartan connection over $M$. A nicely succinct summary is provided by Sharpe:~\cite{Sharpe}
\begin{quote}
A Cartan geometry on $M$ consists of a pair $(P, \om)$, where $P$ is a principal bundle $H \to P \to M$ and $\om$, the Cartan connection, is a differential form on $P$. The bundle generalizes the bundle $H \to G \to G/H$ associated to the Klein setting, and the form $\om$ generalizes the Maurer-Cartan form $\om_G$ on the Lie group $G$. In fact, the \emph{curvature} of the Cartan geometry, defined as $\Om = d \om + \ha [\om, \om]$, is the complete local obstruction to $P$  being a Lie group.

The manifold $P$ may be regarded as some sort of ``lumpy Lie group'' that is homogeneous in the $H$ direction. Moreover, $\om$ may be regarded as a ``lumpy'' version of the Maurer-Cartan form. \emph{The Cartan connection}, $\om$, restricts to the Maurer-Cartan form on the fibers and hence satisfies the structural equation in the fiber directions; but when $\Om \ne 0$ we lose the ``rigidity'' that would otherwise have been provided by the structural equation in the base directions and that would have as a consequence that, locally, $P$ would be a Lie group with $\om$ its Maurer-Cartan form. Thus, the curvature measures this loss of rigidity.
\end{quote}

Since Ehresmannian geometry embraces the principle of geometric naturalness (everything described via maps between vector fields over manifolds), it is worth considering Cartan geometry from this natural point of view. In this context the principal total space of a Lie group factors as $P_{\cal G}={\cal G} \times {\cal G} \simeq G/H \times {\cal H} \times {\cal G}$ and deforms to $\tilde{P_{\cal G}}= \tilde{\cal G} \times {\cal G} \simeq M \times {\cal H} \times {\cal G}$, with the Ehresmann-Maurer-Cartan connection (\ref{EMCc}) deforming to an \emph{Ehresmann-Cartan connection}, $\ve{\f{\cal C} \p{.}} \! \!$, over $\tilde{P_{\cal G}}$, mapping vectors tangent to $\tilde{\cal G}$ to vectors tangent to ${\cal G}$. The Ehresmann-Cartan connection relates to the Cartan connection by $ \ve{\f{\cal C} \p{.}} \! \! (x,y) \, \f{\Th}(y) = \f{C}(x)$. It is good to be able to describe Cartan geometry this way, using maps between vector fields, but working with the Cartan connection over the base is usually more convenient.

Although a Cartan geometry is a deforming Lie group, $\tilde{\cal G}$, which becomes the total space of a principal $H$-bundle, it also has associated $G/H$ frame fibers. To fully describe a Cartan geometry, we may combine these $G/H$ and $H$ fibers, producing a principal $G$-bundle. In this construction, generalizing MacDowell and Mansouri's formulation, a Cartan connection may be identified as a principal $G$-bundle connection, now with a new understanding of why this $G$-connection has $H$-connection and $Lie(G/H)$-valued frame parts.

Integration of functions over a deforming Lie group described as a Cartan geometry can be separated, similarly to (\ref{sepi}), as
\beq
\int_{\tilde{G}} \nf{d^{\tilde{G}} \! z} \, f(z) = \int_{M} \nf{d^{M} \! x} \, \int_H \nf{d^H \! y} \, f ( x,y) \label{sepic}
\eeq
in which $\nf{d^{\tilde{G}} \! z}$ is the measure on ${\tilde{G}}$, $\nf{d^H \! y}$ is the Haar measure on $H$, $\nf{d^{M} \! x}$ is the measure on $M$ provided by the frame part of the Cartan connection, $\f{E}$, multiplied by a constant scale factor, and $f(z) = f(x,y)$ is the integrand as a function of deforming Lie group points, $z$, coordinatized by $x$ over $M$ and $y$ over $H$.

\section{Spin(1,4) deformations}

To describe spacetime and gravity, we start with a Klein geometry using the Lie group $G=Spin(1,4)$ and a $H=Spin(1,3)$ subgroup, producing a four-dimensional homogeneous space, $G/H=Spin(1,4)/Spin(1,3)$, which, we will see, can be identified as de Sitter spacetime. In detail, we choose a $spin(1,3)$ subalgebra of $spin(1,4)$ spanned by six $\ga_{\mu \nu}$ generators, with the complementary vector space, $Lie(Spin(1,4)/Spin(1,3))$, spanned by the four $\ga_{\mu 4}$. As a nice coset representative section we choose
\begin{eqnarray}
r(x) &=& e^{\al s^\pi N_\pi} e^{\al t N_0}  \label{section} \\
 &=& \lp 1 + \al s^\pi N_\pi \rp \lp \cosh(\fr{\al t}{2})+ 2 N_0 \sinh(\fr{\al t}{2}) \rp \notag \\
 &=& \cosh(\fr{\al t}{2}) + 2 N_0 \sinh(\fr{\al t}{2}) + \al e^{\frac{\al t}{2}} s^\pi N_\pi \notag
\end{eqnarray}
which we've calculated out in closed form as $Spin(1,4)$ Lie group elements of the $Cl(1,4)^*$ Clifford group via exponentiation and Clifford multiplication, using space and time coordinates $s^\pi = x^\pi (L/c)$ and $t = x^0 T$, expansion parameter $\al$, and the null generators $N_0 = \ha \ga_{04}$ and $N_\pi = \ha (\ga_{\pi 4} - \ga_{0 \pi})$, and their identities (\ref{nident}). The Maurer-Cartan connection on this section is
\begin{eqnarray}
\f{\Th}(x) &=& r^- \f{d} r \notag \\
 &=& \lp e^{-\al t N_0} e^{-\al s^\pi N_\pi} \rp \lp \f{dt} \lp e^{\al s^\pi N_\pi} \al N_0 e^{\al t N_0} \rp + \f{ds^\pi} \lp \al N_\pi e^{\al s^\pi N_\pi} e^{\al t N_0} \rp \rp \notag \\
 &=& \f{dt} \, \al N_0 + \f{d s^\pi} \al e^{\al t} N_\pi \notag \\
 &=&  \ha \f{\om}^{\mbox{\tiny S}} + \f{E}^{\mbox{\tiny S}} \label{dSMC}
\end{eqnarray}
which we see is the same as the de Sitter connection (\ref{dsc}), $\f{\Th}=\f{H}^{\mbox{\tiny S}}$, with the de Sitter frame and $spin(1,3)$-valued spin connection identified as
\begin{eqnarray}
\f{E}^{\mbox{\tiny S}} &=&
\f{dt} \, \fr{\al}{2} \ga_{04} + \f{d s^\pi} \fr{\al}{2} e^{\al t} \ga_{\pi 4}
= \lp \f{dt} \ga_0 + \f{d s^\pi} e^{\al t} \ga_\pi \rp  \lp \fr{\al}{2} \ga_4 \rp
= \f{e}^{\mbox{\tiny S}} \ph_{\mbox{\tiny S}} \\
\f{\om}^{\mbox{\tiny S}} &=& - \f{d s^\pi} \al e^{\al t} \ga_{0 \pi}
\end{eqnarray}
The fact that there is a representative section of $Spin(1,4)$ such that the Maurer-Cartan form on this section is the de Sitter connection explains why the curvature of the de Sitter connection vanishes. The de Sitter connection is the Maurer-Cartan connection of a Klein geometry with a homogeneous space that is four-dimensional de Sitter spacetime. The spacetime frame part of the de Sitter connection is the frame part of the Maurer-Cartan connection on the homogeneous space. As a homogeneous space, de Sitter spacetime is symmetric under $Spin(1,4)$ transformations.

When we allow this Klein geometry to vary, becoming a Cartan geometry, $S\tilde{pin}(1,4)$, the curvature can become non-zero, describing excitations of the $Spin(1,4)$ Lie group that maintain the integrity of a $Spin(1,3)$ subgroup. The resulting Cartan connection consists of the spin connection and frame,
\beq
\f{C}(x) =  \ha \f{\om} + \f{E} \;\;\;\;\;\;\;\;\;\;\;\;\;\;\;\; \f{\om} = \ha \f{\om}^{\mu \nu} (x) \ga_{\mu \nu} \;\;\;\;\;\;\;\;\;\;\;\;\;\;\;\; \f{E} = \f{e}^\mu (x) \fr{\al}{2} \ga_{\mu 4} \label{ccon}
\eeq
valued in $Lie(Spin(1,3))$ and $Lie(Spin(1,4)/Spin(1,3))$ parts of $Lie(Spin(1,4))$. This Cartan connection, defined over a four-dimensional base manifold, $M$, describes the geometry of spacetime embedded in the deformed $Spin(1,4)$ Lie group, $S\tilde{pin}(1,4)$, just as the section (\ref{section}) embeds de Sitter spacetime in the $Spin(1,4)$ Lie group of the Klein geometry. In this model, physical spacetime is the collection of gauge-related sections of $S\tilde{pin}(1,4)$, with the deforming Lie group and local spacetime geometry described by the spin connection and frame parts of the Cartan connection and its curvature over the base, $M$. In this way, Cartan geometry solves the mystery of how and why the frame part of the $Spin(1,4)$ MacDowell-Mansouri connection is a spacetime frame---it is the frame part of the Cartan connection evaluated at a spacetime section of $S\tilde{pin}(1,4)$.

From the Cartan connection (\ref{ccon}) over $M$, the Ehresmann-Cartan connection form (\ref{ECc}) over all of $S\tilde{pin}(1,4) \simeq M \times Spin(1,3)$ is
\beq
\f{\cal C}(x,y) = \lp h^-(y) \ha \f{\om}(x) h(y) + h^-(y) \f{d^y} h(y) \rp +  h^-(y) \f{E}(x) h(y) \label{eCc}
\eeq
in which $h(y) \in Spin(1,3)$. This connection describes how the entire Lie group manifold deforms to accommodate curved spacetime. The curvature (\ref{Fcov}) of this Cartan geometry, $\ff{\cal F}(x,y) = h^-(y) \ff{F}(x) h(y)$, is described by the curvature of the Cartan connection over $M$,
\beq
\ff{F}(x) = \f{d} \f{C} + \f{C} \f{C} = (\ha \ff{R} + \fr{\al^2}{4} \f{e} \f{e}) + \ff{T} \fr{\al}{2} \ga_4 \label{ccurv}
\eeq
with $\ff{R}$ the Riemann curvature (\ref{R}) and $\ff{T}$ the torsion (\ref{T}).

Other representative sections of $S\tilde{pin}(1,4)$ obtainable via $Spin(1,3)$ gauge transformations correspond to gauge-equivalent spacetimes. The decomposition of $Spin(1,4)$, and of $S\tilde{pin}(1,4)$, into fibers and sections presumes that each fiber intersects a section once and only once. If this is not true, as is sometimes the case for arbitrary Lie subgroup fibers, $H \subset \tilde{G}$, and sections, $\si$, the decomposition necessarily becomes more elaborate. There are two ways this can happen. One possibility is that there are $H$ fibers in $\tilde{G}$ that do not intersect $\si$. The second possibility is that there are $H$ fibers that intersect $\si$ more than once. In the first case, we can choose a larger section intersected by these fibers. In the second case, we might be able to choose a different section to avoid multiple intersection, or we might have to deal with this \emph{Gribov ambiguity} in some other way.

\section{Lie group decompositions and the de Sitter subgroup}

The topology of Lie group manifold embeddings and deformations is complex and fascinating~\cite{Randono, Randono2}, but will largely be left to consider in future work. We will, though, briefly touch on some topological issues. First, we should admit to having used some unusual conventions and abuses of language.

We have chosen to work with spin groups rather than orthogonal groups because Clifford algebra is efficient for computations and because we eventually need to deal with spinors to describe fermions. We follow the convention that $Spin(1,q)$ is the simply connected double cover, often labeled $Spin_0(1,q)$, of the identity component, $SO_0(1,q)$, of the orthogonal group, $O(1,q)$. What we have been calling ``de Sitter spacetime,'' the coset space $Spin(1,4)/Spin(1,3)$, is half of what is usually called ``de Sitter space,'' the coset space $O(1,4)/O(1,3)$. De Sitter space can also be described as a hyberboloid embedded in five-dimensional Minkowski space, defined by
$$
(z^0)^2 - \sum_{a=1}^4 (z^a)^2 = -\frac{1}{\al^2}  
$$
Half of this hyperboloid corresponds to de Sitter spacetime,
\begin{eqnarray*}
z^0 &=& \fr{1}{\al} \sinh(\al t) + \fr{\al}{2} R^2 e^{\al t} \\
z^\pi &=& e^{\al t} s^\pi \\
z^4 &=& \fr{1}{\al} \cosh(\al t) - \fr{\al}{2} R^2 e^{\al t}
\end{eqnarray*}
with $t$ and $s^\pi$ ranging from minus to plus infinity (in temporal units) and $R^2 = \sum_{\pi=1}^3 (s^\pi)^2$. The other half of de Sitter space, the \emph{dual de Sitter spacetime}, is obtained by mirroring de Sitter spacetime through the origin of five-dimensional Minkowski space. (Our de Sitter spacetime has also been called ``elliptic de Sitter space,'' $dS/\mathbb{Z}_2$~\cite{dS}.) Although the embedding of de Sitter spacetime in five-dimensional Minkowski space can be useful, we consider its embedding in $Spin(1,4)$ to be more fundamental. The de Sitter spacetime also provides a more physically realistic cosmological model than the full de Sitter space hyperboloid, describing an eternally exponentially expanding universe.

The chosen representative section (\ref{section}) of $Spin(1,4)$ corresponding to de Sitter spacetime,
\beq
r(s,t) = n(s) \, a(t) = e^{\al s^\pi N_\pi} e^{\al t N_0} = e^{\al t N_0} + \al e^{\frac{\al t}{2}} s^\pi N_\pi \label{rsec}
\eeq
comes from an \emph{Iwasawa-like decomposition}, by which some elements of $Spin(1,4)$ can be factored as
\beq
g(s,t,b) = n(s) \, a(t) \, h(b) = \pm e^{\al s^\pi N_\pi} e^{\al t N_0} e^{\frac{1}{2} b^{\mu \nu} \ga_{\mu \nu}} \label{decom}
\eeq
in which $n(s) = e^{\al s^\pi N_\pi}$ is the exponentiation of null generators (\ref{tngen}), $a(t) = e^{\al t N_0}$ is the exponentiation of the temporal generator, and $h(b) = \pm e^{\frac{1}{2} b^{\mu \nu} \ga_{\mu \nu}}$ are $Spin(1,3)$ subgroup elements.\footnote{Some elements of $Spin(1,3)$ can't be reached by exponentiating bivectors, requiring the minus~\cite{Lounesto}.} Since $Spin(1,3)$ is not compact, this is not a usual Iwasawa decomposition, but follows similar lines. The null generators, $N_\pi$, are the \emph{negative root vectors} with respect to a Cartan-Weyl decomposition of $Lie(Spin(1,4))$ having $N_0$ in the Cartan subalgebra. The temporal and three null de Sitter generators exponentiate to produce the \emph{de Sitter subgroup}. This four-dimensional solvable Lie subgroup is embedded in $Spin(1,4)$ as the section $g(s,t,0) = r(s,t)$. As a Lie group, the product of any two de Sitter spacetime points is a third point,
\begin{eqnarray}
r(s,t) \, r(s',t') &=& \lp e^{\al t N_0} + \al e^{\frac{\al t}{2}} s^\pi N_\pi \rp  \lp e^{\al t' N_0} + \al e^{\frac{\al t'}{2}} s'^\rh N_\rh \rp \notag \\
&=& e^{\al \lp t + t' \rp N_0} +  \al e^{\frac{\al \lp t + t' \rp}{2}} \lp s^\pi + e^{-\al t} s'^\pi \rp N_\pi \notag \\
&=& r( s + e^{-\al t} s', t+t') \notag
\end{eqnarray}
with the identity at $r(0,0)$ and inverses given by $r^-(s,t) = r( -e^{\al t} s, -t)$. This description of de Sitter spacetime as a Lie group (sometimes going by the name $AN(3)$) is not new but is not widely appreciated.

Along with the embedded de Sitter spacetime obtained from exponentiating $N_\mu$, there is an algebraically \emph{dual de Sitter spacetime} embedded in $Spin(1,4)$. This dual de Sitter spacetime, with representative section
$$
r'(s,t) = n'(s) \, a(t) = e^{\al s^\pi N'_\pi} e^{\al t N_0} = e^{\al t N_0} + \al e^{\frac{- \al t}{2}} s^\pi N'_\pi
$$
is generated by $N_0$ and the three \emph{positive root vectors},
$$
N'_\pi = \ha \lp \ga_{\pi 4} + \ga_{0 \pi} \rp 
$$
dual to $N_\pi$. The Cartan involution, $N_\pi \leftrightarrow N'_\pi$, corresponds to a duality outer-automorphism of the $spin(1,4)$ Lie algebra, with this automorphism relating two \emph{regions} of the Lie group. To cover $Spin(1,4)$, we can use $Spin(1,3)$ gauge transformations of $r$, as in (\ref{decom}), and of $r'$. If we consider the $Spin(1,4)$ outer-automorphism as a gauge transformation, then deformations of $Spin(1,4)$ to $S\tilde{pin}(1,4)$ can be described by a Cartan connection over $M$ modeled on de Sitter spacetime with representative, $r$, in one region, with gauge transformations giving the Ehresmann-Cartan connection form over the entire deforming Lie group.

The fact that our chosen representative section (\ref{rsec}), $r$, is a Lie subgroup of $Spin(1,4)$ has interesting implications. One implication is that this de Sitter subgroup will also be a Lie subgroup of any group containing $Spin(1,4)$, such as $Spin(12,4)$ or $E_{8(-24)}$. Another implication is that analysis over de Sitter spacetime will relate to representations of the de Sitter subgroup. However, it is important to keep in mind that even though our chosen section (\ref{rsec}) is a Lie subgroup, other, gauge-equivalent sections of $Spin(1,4)$ also describe de Sitter spacetime without being subgroups. Also, the de Sitter subgroup, as a Lie group, cannot be considered, alone, as de Sitter spacetime because its Killing form, and thus its metric, is degenerate. The de Sitter subgroup can only be considered de Sitter spacetime as a representative of the coset space $Spin(1,4)/Spin(1,3)$ in $Spin(1,4)$, in which the Maurer-Cartan form separates into a frame and spin connection, and the non-degenerate metric is inherited from $Spin(1,4)$.

\section{Deformation dynamics}

To prescribe dynamics for $S\tilde{pin}(1,4)$ and larger deforming Lie groups, we can specify an action dependent on the Cartan connection. Following our philosophical desire for geometric unity~\cite{Eric}, we prefer an action functional that is an integral over the total space of the deforming Lie group. The most compelling possibility is the Yang-Mills action,
$$
S = \fr{1}{2} \int_{S\tilde{pin}(1,4)} \lp \ff{\cal F}(z) , \nf{\accentset{\cal C}{\star} \cal F} \rp
 = \fr{1}{4} \int_{S\tilde{pin}(1,4)} \nf{d^{S\tilde{pin}(1,4)} \! z} \;  g^{ac} g^{bd} n_{AB} {\cal F}^A_{ab} {\cal F}^B_{cd}
$$
using the $spin(1,4)$ Killing form, $n_{AB}$, the curvature, $\ff{\cal F}$, of the Cartan geometry, and the Hodge duality operator, $\accentset{\cal C}{\star}$, and metric, $g^{ab}(z)$, derived from the Ehresmann-Cartan connection form (\ref{ECc}), much as the Hodge duality operator and metric over the Lie group manifold were constructed from the Ehresmann-Maurer-Cartan connection form (\ref{metric}). Because the deforming Lie group maintains the integrity of a $Spin(1,3)$ subgroup, the curvature is gauge-covariant (\ref{Fcov}) under the chosen $Spin(1,3)$, and is horizontal over sections (a $2$-form on $M$). Thus, the integrand is $Spin(1,3)$ invariant and the action separates (\ref{sepic}) and integrates to
$$
S = \fr{1}{4} \int_M \nf{d^M \! x} \, \int_{Spin(1,3)} \nf{d^{Spin(1,3)} \! y} \;  \fr{16}{\al^4} g^{ik}(x) g^{jm} n_{AB} F_{ij}{}^A(x) F_{km}{}^B = \fr{8 V}{\al^4} \int_M \lp \ff{F}(x) , \ff{\star F} \rp 
$$
with $V$ the volume of the $Spin(1,3)$ Lie group manifold. In the resulting integral, $\ff{F}$ is the curvature of the Cartan connection (\ref{ccurv}),
$$
g^{ij}(x) = e_\mu{}^i \et^{\mu \nu} e_\nu{}^j
$$
is the metric inverse, and $\star$ is the Hodge duality operator over $M$, both using the frame part, $\f{E} = \f{dx^i} e_i{}^\mu \fr{\al}{2} \ga_{\mu 4}$, of the Cartan connection (\ref{ccon}). It is very nice that the Hodge duality operator and Yang-Mills action integrated over the entire deforming Lie group reduces to the spacetime Hodge and Yang-Mills action for the Cartan connection integrated over a spacetime section. Separating the curvature into parts (\ref{ccurv}) and using the Hodge area identity (\ref{eeg}), the action becomes
$$
S = -\fr{2 V}{\al^2} \int_M \nf{d^4  x} \left| e \right| \left\{ R + 6 \al^2 + \fr{1}{4 \al^2} R_{ij}{}^{\mu \nu} R^{ij}{}_{\mu \nu} + \fr{1}{4} T_{ij}{}^{\mu} T^{ij}{}_{\mu} \right\}
$$
which includes an Einstein-Hilbert term, a positive cosmological constant, a Kretschmann scalar term, $R_{ij}{}^{\mu \nu} R^{ij}{}_{\mu \nu}$, (also known as a Stephenson-Kilmister-Yang (SKY) term), and a torsion term---providing a reasonable action for gravity.

Although the expressions for the Cartan connection and the MacDowell-Mansouri connection are identical, the geometric model of Cartan gravity, with $S\tilde{pin}(1,4)$ deforming as described by its Cartan connection, is more succinct and elegant than the MacDowell-Mansouri framework of a $Spin(1,4)$ bundle with connection over a four dimensional base. Philosophically, it is more satisfying to begin with a unified geometric structure, such as the $Spin(1,4)$ Lie group, and consider its deformation, than to begin with a more disjoint structure, such as a principal and associated fiber bundle. Also, we have now found a satisfactory answer to why the frame part of the MacDowell-Mansouri connection is related to spacetime. In Cartan geometry, the frame is the part of the Cartan connection related to spacetime subspaces of the deforming Lie group. Instead of a symmetry breaking process, such as was required in the MacDowell-Mansouri formulation, Cartan geometry employs a \emph{symmetry keeping} process. Maintaining the rigidity of $Spin(1,3)$ fibers within $S\tilde{pin}(1,4)$ determines how the $Spin(1,4)$ symmetry breaks, with spacetime described as the curving, broken $Spin(1,4)/Spin(1,3)$ part of the geometry. The fibers and base spacetime of Cartan geometry arise naturally from the original Lie group geometry and deformations preserving a subgroup.

Given our success in describing spacetime using $Spin(1,4)$ Klein geometry and Cartan geometry, we may naturally wonder whether this same framework will work with larger Lie groups, such as $Spin(12,4)$, to describe the gauge fields of the Standard Model and Grand Unified Theories as well as gravity; however, a problem immediately arises. Even when we choose the largest possible subgroup, the homogenous space of the resulting Klein geometry is typically very high-dimensional, such as fifteen-dimensional for $Spin(12,4)/Spin(12,3)$. This is bad for modeling four-dimensional spacetime. We could employ Kaluza-Klein theory to compactify some of these extra spacetime dimensions, but that would complicate matters and raise other issues. So it appears, at first, that Cartan geometry does not obviously give a unified GraviGUT. But what if we generalize Cartan geometry and consider deformations of a large Lie group, such as $Spin(12,4)$, with four-dimensional submanifolds corresponding to spacetime? There is no reason we can't choose a $Spin(1,4)$ subroup of $Spin(12,4)$ and model four-dimensional spacetime on a representative subspace of $Spin(12,4)$ corresponding to $Spin(1,4)/Spin(1,3)$ de Sitter spacetime within the $Spin(1,4)$ subgroup.

\section{Generalized Cartan connections}

How can we generalize Cartan geometry to describe a large Lie group deforming over cosets of a subgroup? If we start with a Lie group, $G$, containing a subgroup, $G'$, which contains a subgroup, $H$, then we can allow $G'/H$ to become wavy while maintaining $H$, as in Cartan geometry, and also maintaining the rigidity of $G/G'$. If we wish to integrate over the large Lie group, we can use (\ref{sepi}) twice to separate the integral as
\beq
\int_G \nf{d^G \! z} \, f(z) = \int_{G/G'} \nf{d^{G/G'} \! w} \, \int_{G'/H} \nf{d^{G'/H} \! x} \, \int_H \nf{d^H \! y} \, f (w, x, y) \label{sepi2}
\eeq
If we allow $G'/H$ to become wavy, calling it $M$, and provided $f$ is independent of $w$ and $y$, the integral over the deforming Lie group becomes
\beq
\int_{\tilde{G}} \nf{d^{\tilde{G}} \! z} \, f(x) = V_{G/G'} \, V_H \, \int_M \nf{d^M \! x} \, f(x) \label{sepi3}
\eeq
with $V_{G/G'}$ and $V_H$ the volumes of $G/G'$ and $H$. This worked when the deformation was described by a Cartan connection (\ref{Carcon}) valued in $Lie(G') = Lie(H) + Lie(G'/H)$, and it also works if we allow more general excitations, described by a \emph{generalized Cartan connection},
\beq
\f{C}(x) = \ha \f{\om} + \f{E} + \f{A} \label{gCc}
\eeq
in which $\f{\om}$, $\f{E}$, and $\f{A}$ are $1$-forms over $M$ valued in $Lie(H)$, $Lie(G'/H)$, and $Lie(G/G')$. This generalized Cartan connection, valued in $Lie(G)$, describes how the large Lie group deforms and twists over $M$. It is the pullback of the \emph{generalized Ehresmann-Cartan connection form} over $\tilde{G}$,
\beq
\f{\cal{C}}(w,x,y) = r^-(w) h^-(y) \f{C}(x) h(y) r(w) + r^-(w) h^-(y) \, \f{d} \, h(y) r(w) \label{gECg} 
\eeq
in which $r(w)$ is a representative element of $G/G'$, and $h(y)$ is an element of $H$. The curvature of this generalized Cartan geometry is
\beq
\ff{\cal F}(w,x,y) = \f{d} \f{\cal C} + \f{\cal C} \f{\cal C} = r^-(w) h^-(y) \ff{F}(x) h(y) r(w) \label{gFcov}
\eeq
in which
\begin{eqnarray}
\ff{F}(x) & = & \f{d} \f{C} + \f{C} \f{C} \notag \\
&=& \ha \lp \f{d} \f{\om} + \ha \f{\om} \f{\om} + \f{E} \f{E} \rp + \lp \f{d} \f{E} + \ha \f{\om} \f{E} + \ha \f{E} \f{\om} + \f{A} \f{E} + \f{E} \f{A} \rp + \lp \f{d} \f{A} + \f{A} \f{A} \rp \notag \\
&=& \lp \ha \ff{R} + \f{E} \f{E} \rp + \f{D} \f{E} + \ff{F}_A
\label{gCccurv}
\end{eqnarray}
is the curvature of the generalized Cartan connection over $M$.

\section{Deforming Lie group regions}

The generalized Cartan geometry picture is very succinct: a sheaf of four-dimensional spacetimes modeled on de Sitter spacetime is embedded in a large deforming Lie group, with Lorentz and other gauge transformations relating spacetime subspaces. However, as we encountered in the decomposition of $Spin(1,4)$, not all regions of a deforming Lie group can be reached along fibers (orbits of Lorentz or usual gauge transformations) over embedded spacetime. For many Lie groups there will be regions not reachable by gauge orbits intersecting a chosen spacetime. In the example of $Spin(1,4)$, a duality automorphism was employed as a \emph{large gauge transformation}, allowing sheaves of gauge-related spacetimes to cover the Lie group. For a more complicated example, consider the Lie group $G=Spin(4,4)$. We can choose a $G'=Spin(1,4)$ and $H=Spin(1,3)$ subgroup, with a representative of de Sitter spacetime, $G'/H$, embedded in $Spin(1,4)$ and therefore embedded in $Spin(4,4)$. The sheaf of gauge-transformed spacetimes generated by $Lie(G/G')$ and $spin(1,3)$, corresponding to some inner-automorphisms of $Spin(4,4)$, will not completely cover $Spin(4,4)$, nor will sheaves related by duality. This is because $Spin(4,4)$ also has a \emph{triality} outer-automorphism, transforming embedded de Sitter spacetime to two other de Sitter spacetime copies embedded in $Spin(4,4)$. Large gauge transformations, relating these three spacetimes, cannot be described via usual Lorentz or gauge transformations of the connection (\ref{gECg}). How, then, should we think of embedded spacetime in this ambiguous case? Is physical spacetime one of these triality-related spacetimes, or somehow a superposition of all three? A similar problem, \emph{Gribov ambiguity}, is well known in gauge theory, and the usual solution is to integrate over multiple Gribov regions, matching at the boundaries. This same approach works within our generalized Cartan geometry, in which we must embed multiple spacetimes and their corresponding fiber bundles to properly cover and describe the deforming Lie group. This approach is unusual because it implies we have several equivalent spacetimes, but is similar to having many equivalent spacetime sections of a fiber bundle. Another way to think of this geometry, from a top-down perspective, is that a deforming Lie group can have regions described by different sheaves of spacetime, with different matter content, and different but compatible connections.

In the example of $Spin(4,4)$, a \emph{triality automorphism} induces an automorphism of the Lie algebra, $T : spin(4,4) \to spin(4,4)$, with $T^3$ the identity. If we consider an embedded de Sitter spacetime representative, $S_{\mbox{\tiny I}}$, generated by some $N_\mu$, as in (\ref{tngen}) and (\ref{section}), then two other copies of de Sitter spacetime, $S_{\mbox{\tiny II}}$ and $S_{\mbox{\tiny III}}$, in different regions of $Spin(4,4)$, with non-trivial intersection, are generated by $T \, N_\mu$ and $T^2 \, N_\mu$. For example, three sets of triality-related de Sitter generators could be
$$
\begin{array}{rcl}
N_0 &=& \ha \ga_{04}                             \\
N_1 &=& \ha \lp \ga_{14} - \ga_{01} \rp  \\
N_2 &=& \ha \lp \ga_{24} - \ga_{02} \rp  \\
N_3 &=& \ha \lp \ga_{34} - \ga_{03} \rp  \\
\end{array}
$$
$$
\begin{array}{rcl}
T N_0 &=& \fr{1}{4} \lp -\ga_{15} - \ga_{26} - \ga_{37} - \ga_{04} \rp \\
T N_1 &=& \fr{1}{4} \lp \ga_{14}  + \ga_{23}  + \ga_{67} + \ga_{05} \rp - \fr{1}{4} \lp -\ga_{01}  - \ga_{27} + \ga_{36} + \ga_{45}   \rp \\
T N_2 &=& \fr{1}{4} \lp -\ga_{13} + \ga_{24} - \ga_{57} + \ga_{06}  \rp - \fr{1}{4} \lp -\ga_{02} + \ga_{17} - \ga_{35} + \ga_{46}  \rp \\ 
T N_3 &=& \fr{1}{4} \lp \ga_{12} + \ga_{34} + \ga_{56} + \ga_{07} \rp - \fr{1}{4} \lp -\ga_{03} - \ga_{16} + \ga_{25}  + \ga_{47}  \rp \\
\end{array}
$$
$$
\begin{array}{rcl}
T^2 N_0 &=& \fr{1}{4} \lp \ga_{15} + \ga_{26} + \ga_{37} - \ga_{04} \rp \\
T^2 N_1 &=& \fr{1}{4} \lp \ga_{14}  + \ga_{23} + \ga_{67} - \ga_{05} \rp - \fr{1}{4} \lp -\ga_{01} + \ga_{27} - \ga_{36} - \ga_{45} \rp \\
T^2 N_2 &=& \fr{1}{4} \lp -\ga_{13}  + \ga_{24} - \ga_{57} - \ga_{06} \rp - \fr{1}{4} \lp -\ga_{02} - \ga_{17} + \ga_{35} - \ga_{46}  \rp \\
T^2 N_3 &=& \fr{1}{4} \lp \ga_{12}  + \ga_{34} + \ga_{56} - \ga_{07} \rp - \fr{1}{4} \lp -\ga_{03} + \ga_{16} - \ga_{25} - \ga_{47}  \rp \\
\end{array}
$$
These generators reside in three different \emph{Lie algebra regions}, and each set of four exponentiates to give a different de Sitter spacetime representative residing in a different Lie group region. Each of the three copies of de Sitter spacetime will have a corresponding frame, $\f{e}^{\mbox{\tiny S}}_{\mbox{\tiny I,II,III}}$, Higgs vector, $\ph^{\mbox{\tiny S}}_{\mbox{\tiny I,II,III}}$, and spin connection, $\f{\om}^{\mbox{\tiny S}}_{\mbox{\tiny I,II,III}}$. All of $Spin(4,4)$ is reachable via Lorentz or gauge transformations from these three copies of embedded spacetime, or from their duals. The single spacetime that we perceive can be described by the generalized Ehresmann-Cartan connection form, with connections in different regions related by automorphisms, similar to how gauge fields in spacetime are described by connections related by inner-automorphisms over the total space of a principal bundle. 

Another way of understanding the existence of three spacetimes in $Spin(4,4)$ is the following: Choose a $Spin(1,4)$ (or similar) subgroup of $Spin(4,4)$. Decompose that $Spin(1,4)$ into $Spin(1,3)$ and the four-dimensional coset space. The pullback of $Spin(4,4)$'s Maurer-Cartan form onto a representative of that coset space will be the combined spin connection and frame of de Sitter spacetime (\ref{dSMC}). There are two other partially overlapping $Spin(1,4)$'s in $Spin(4,4)$ related to the first by triality automorphism. By that automorphism, the spin connection and frame coefficients for each of those triality-related spacetimes, from the pullback of the $Spin(4,4)$ Maurer-Cartan form, must be identical. Therefore, we have three de Sitter spacetimes, which we identify as vacuum spacetimes in three different regions. If we wish, we can consider independent or identical perturbations of the spin connections and frames of those spacetimes.

Lie group excitations in various regions might not be perfectly related by large gauge transformations, in which case it may be necessary to describe the deforming Lie group with independent generalized Cartan connections, such as $\f{C}_{\mbox{\tiny I}}$, $\f{C}_{\mbox{\tiny II}}$, and $\f{C}_{\mbox{\tiny III}}$, over the different regions. The spin connection, $\f{\om}$, and frame-Higgs, $\f{E}$, of physical spacetime will then be identified with some superposition of $\f{\om}_{\mbox{\tiny I}}$, $\f{\om}_{\mbox{\tiny II}}$, or $\f{\om}_{\mbox{\tiny III}}$, and $\f{E}_{\mbox{\tiny I}}$, $\f{E}_{\mbox{\tiny II}}$, or $\f{E}_{\mbox{\tiny III}}$. Although their field components may differ, the Lie algebraic generators of these fields must be related by triality, such as in $\f{\om}_{\mbox{\tiny I}} = \ha \f{\om}_{\mbox{\tiny I}}^{\mu \nu} \ga_{\mu \nu}$ and $\f{\om}_{\mbox{\tiny II}} = \ha \f{\om}_{\mbox{\tiny II}}^{\mu \nu} T \, \ga_{\mu \nu}$. These may be combined in a \emph{superposed Cartan connection},
\beq
\f{C} = \f{C}_{\mbox{\tiny I}} + \f{C}_{\mbox{\tiny II}} + \f{C}_{\mbox{\tiny III}} = \ha \f{\om}_{\mbox{\tiny I}} + \ha \f{\om}_{\mbox{\tiny II}} + \ha \f{\om}_{\mbox{\tiny III}} + \f{E}_{\mbox{\tiny I}} + \f{E}_{\mbox{\tiny II}}+ \f{E}_{\mbox{\tiny III}} + \f{A}_{\mbox{\tiny I}} + \f{A}_{\mbox{\tiny II}} + \f{A}_{\mbox{\tiny III}} \label{gCct}
\eeq
in which the generators of the Roman numbered fields, but not necessarily their coefficients, are related by a triality automorphism. The \emph{regional Cartan connections}, $\f{C}_{\mbox{\tiny I,II,III}}$, may be valued in overlapping parts of the Lie algebra---with some of their generators invariant under triality. We may express the Roman numbered fields as triality maps of Arabic numbered fields, such as
$$
\f{\om}_{\mbox{\tiny I}} = \ha \f{\om}_{\mbox{\tiny I}}^{\mu \nu} \ga_{\mu \nu} = \f{\om}_1 \;\;\;\;\;\;\;\;\;\;\; 
\f{\om}_{\mbox{\tiny II}} = \ha \f{\om}_{\mbox{\tiny II}}^{\mu \nu} T \ga_{\mu \nu} = T \f{\om}_2 \;\;\;\;\;\;\;\;\;\;\; 
\f{\om}_{\mbox{\tiny III}} = \ha \f{\om}_{\mbox{\tiny III}}^{\mu \nu} T^2 \ga_{\mu \nu} = T^2 \f{\om}_3
$$
To ensure that the superposed Cartan connection describes our one spacetime, we presume that the spin connections, gravitational frames, and gauge fields in different regions have equal coefficients, so
$$
\f{\om}_1 = \f{\om}_2 = \f{\om}_3 = \f{\om} \;\;\;\;\;\;\;\;\;\;\;
\f{e}_1^\mu = \f{e}_2^\mu = \f{e}_3^\mu = \f{e}^\mu \;\;\;\;\;\;\;\;\;\;\;
\f{A}_1 = \f{A}_2 = \f{A}_3 = \f{A}
$$
If we wish to maintain perfect symmetry under triality, we would also presume that the Higgs vacuum expectation values (\ref{vev}) are equal in the different regions,
$$
\ph^4_0{\mbox{\tiny I}} = \ph^4_0{\mbox{\tiny II}} = \ph^4_0{\mbox{\tiny III}} =  \ph^4_0 =\frac{\al}{2}
$$
Alternatively, we can allow for the possibility that the Higgs vevs may be different,
$$
\ph^4_0{\mbox{\tiny I}} = \frac{\al_1}{2}  \;\;\;\;\;\;\;\;\;\;\;
\ph^4_0{\mbox{\tiny II}} = \frac{\al_2}{2}  \;\;\;\;\;\;\;\;\;\;\;
\ph^4_0{\mbox{\tiny III}} = \frac{\al_3}{2}
$$
with correspondingly different $\f{E}_{1,2,3}$, such as $\f{E}_2 = \f{e}^\mu \frac{\al_2}{2} \ga_{\mu 4}$.

Although the generalized Cartan connection, and its superposed variant, can describe some excitations of our large Lie group, it is possible that we have not generalized Cartan geometry enough---that we need to accommodate deformations that cannot be described by our connection. For example, we might wish to allow excitations described by a connection $1$-form that is not necessarily a spacetime $1$-form, but is a $1$-form within the deforming Lie group.

\section{Superconnections}

There are a few good ways to further extend the generalized Cartan connection and describe even more general Lie group excitations. One way is to consider excitations in the affine space of connections, using a superconnection consisting of a spacetime $1$-form plus a $1$-form on the space of connections. This is the geometric framework of topological quantum field theory and the BRST formulation of quantum gauge field theory, in which BRST \emph{ghosts} are anti-commuting (fermionic) fields valued in the same Lie algebra as the connection. This approach, which is well-established, has been explored in earlier work, and here we will propose something more unusual. A different way to describe extended Lie group deformations is similar, but more natural and direct: we simply consider the \emph{extended generalized Cartan connection},
\beq
\f{G}(x) = \f{C} + \f{\Psi} \label{egCc}
\eeq
to be a Lie algebra valued $1$-form on the deforming Lie group manifold, consisting of spacetime and non-spacetime $1$-forms, valued in complementary parts of $Lie(G)$, evaluated at points, $x$, of embedded spacetime, $M$. This decomposes into a generalized Cartan connection spacetime $1$-form (\ref{gCc}), $\f{C}$, plus the non-spacetime $1$-form, $\f{\Psi}$, which can be treated as an anti-commuting (fermionic, Grassmann) field, $\ud{\Psi}$, over spacetime. (To describe BRST ghosts, we could allow both $\f{C}$ and $\f{\Psi}$ to be valued in all of $Lie(g)$, and not just in complementary parts.)

The underdot in $\ud{\Ps}$ indicates vertical $1$-form components, orthogonal to spacetime. We have not yet specified the precise directions of the non-spacetime $1$-form, $\f{\Ps}$, within $\tilde{G}$, corresponding to $\ud{\Ps}$. A reasonable choice is that the fermionic $1$-form directions, $\f{\xi}^X$, are Maurer-Cartan form components, dual to the fermionic generator vectors, $\ve{\xi}_X(x)$, so that $\Psi_X{}^Y$ in $\f{\Ps}=\f{\xi}^X \Psi_X{}^Y T_Y$ is diagonal. This implies that we can write $\ud{\Ps}$ using super-components multiplying Lie algebra generators or using components multiplying superalgebra generators,
\beq
\ud{\Ps} = \ud{\Ps}^X Q_X = \ud{\xi}^Y \Psi_Y{}^X Q_X = \ud{\xi}^Y \Psi^X \de_Y^X Q_X = \Ps^X \ud{Q}_X \label{ferm}
\eeq
In this expression, the superalgebra generators, $\ud{Q}_X = \ud{\xi}^X Q_X$ (no sum), act in the Lie bracket as 
$$
\lb \ud{Q}_X, \ud{Q}_Y \rb = \ud{\xi}^X \ud{\xi}^Y \lp c_{XY}{}^A T_A + c_{XY}{}^Z Q_Z \rp 
= \udd{c}_{XY}{}^A T_A + \udd{\od{c}}_{XY}{}^Z \ud{Q}_Z = \lb \ud{Q}_Y, \ud{Q}_X \rb
$$
implying that the Lie bracket of these elements naturally produces a Lie super-bracket, in which $\udd{c}_{XY}{}^A=\udd{c}_{\, Y \! X}{}^A$ and $\udd{\od{c}}_{XY}{}^Z$ are the superstructure constants. Over spacetime, the \emph{superconnection} corresponding to the extended generalized Cartan connection is
\beq
\udf{G}(x) = \f{C} + \ud{\Psi} \label{sc}
\eeq
This superconnection, consisting of (bosonic) $1$-form, $\f{C}$, and anti-commuting (fermionic) parts, $\ud{\Psi}$, describes excitations beyond what is described by a generalized Cartan connection. This kind of deforming Lie group, $\tilde{G}$, described by a superconnection, has also been called a ``soft group manifold''~\cite{Soft}, in the spirit of Salvador Dali's \emph{Soft Self-Portrait}, and an ``almost Lie group''~\cite{Almost}.

Similarly to (\ref{gECg}), the \emph{extended generalized Ehresmann-Cartan connection form} of an excited Lie group is
\beq
\f{\cal{G}}(w,x,y) = r^-(w) h^-(y) \f{G}(x) h(y) r(w) + r^-(w) h^-(y) \, \f{d} \, h(y) r(w) \label{egECcf} 
\eeq
with the extended generalized Cartan connection (\ref{egCc}), $\f{G}(x)$, evaluated at embedded spacetime points. This connection form is well-defined locally near embedded spacetime points, and possibly globally, depending on topology. The \emph{extended generalized Ehresmann-Cartan curvature} is
\beq
\ff{\cal{F}}(w,x,y) = \f{d} \, \f{\cal{G}} + \f{\cal{G}} \, \f{\cal{G}} = r^-(w) h^-(y) \ff{F}(x) h(y) r(w) \label{egECc}
\eeq
in which the \emph{extended generalized Cartan curvature}, evaluated at spacetime points embedded in $\tilde{G}$, is
\beq
\ff{F}(x) = \f{d} \, \f{G}(x) + \ha [ \f{G}, \f{G} ]
= \lp \f{d} \f{C} + \f{C} \f{C} \rp + \lp \f{d} \f{\Ps} + [\f{C}, \f{\Ps}] \rp + \lp \f{\Ps} \f{\Ps} \rp \label{egCcc}
\eeq
This curvature, a $2$-form on the deforming Lie group, corresponds to the \emph{supercurvature},
\beq
\udff{F}(x) = \f{d} \, \udf{G}(x) + \ha [ \udf{G}, \udf{G} ]
= \lp \f{d} \f{C} + \f{C} \f{C} \rp + \lp \f{d} \ud{\Ps} + [\f{C}, \ud{\Ps}] \rp + \lp \ud{\Ps} \ud{\Ps} \rp \label{scurv}
\eeq
of the superconnection (\ref{sc}) over spacetime, $M$. The mixed fermionic and $1$-form part of this supercurvature is the covariant derivative,
$$
\f{D} \ud{\Ps} = \f{d} \ud{\Ps} + [\f{C}, \ud{\Ps}]
$$
in which the fermionic part of the superconnection, $\ud{\Ps}$, necessarily anticommutes with the spacetime $1$-form, $\f{C}$.

The extended generalized Ehresmann-Cartan connection form (\ref{egECcf}) can be used in the usual way to create a Hodge star operator in the deforming Lie group. Using this Hodge star, $\accentset{\cal G}{\star}$, the most natural action for the superconnection integrated over the entire deforming Lie group is an \emph{extended generalized Yang-Mills action},
\beq
S = \fr{1}{2} \int_{\tilde{G}} \lp \ff{\cal F}(z) , \nf{\accentset{\cal G}{\star} \cal F} \rp
 = \fr{1}{4} \int_{\tilde{G}} \nf{d^{\tilde{G}} \! z} \; g^{ac} g^{bd} n_{AB} {\cal F}^A_{ab} {\cal F}^B_{cd} \label{sa1}
\eeq
using the Killing form, $n_{AB}$, the extended generalized Ehresmann-Cartan curvature, $\ff{\cal{F}}$, and the deforming Lie group's metric, $g^{ab}(z)$, from the components of $\f{G}$, as in (\ref{metric}). If the Lie group, $G$, has subgroups $H \subset G' \subset G$, and deforms as described by its extended generalized Ehresmann-Cartan connection form (\ref{egECcf}), while maintaining the integrity of $G/G'$ and $H$, then the extended generalized Ehresmann-Cartan curvature is covariant (\ref{egECc}) and the integrand of (\ref{sa1}) depends only on $x$, the position on an embedded spacetime manifold, $M$, modeled on $G'/H$. If that is the case, the action separates (\ref{sepi3}) and integrates to
\beq
S = V_{G/G'} \, V_H \, \fr{1}{4} \int_M \nf{d^M \! x} \, g^{ac}(x) g^{bd}(x) n_{AB} F^A_{ab}(x) F^B_{cd}(x) \label{sa2}
\eeq
in which the non-spacetime components of the curvature now do not necessarily vanish.

We can use the deforming Lie group metric, $g^{ab}$, along the \emph{fermionic} directions, $g^{xy}$, and along spacetime directions, $g^{ij}$, to define the \emph{super-Hodge star}, $\accentset{.}{\star}$~\cite{sHodge}, and write this action as
\beq
S = \fr{V}{2} \int_M \lp \udff{F}(x) , \nf{\accentset{.}{\star} F} \rp \label{sa3}
\eeq
in which $\udff{F}$ is the supercurvature (\ref{scurv}) and the volumes have been combined into $V = V_{G/G'} \, V_H$. Separating the supercurvature into its components (\ref{scurv}), the action is
\beq
S = \fr{V}{2} \int_M \Big\{ \big( \ff{F}(x) , \nf{\star F} \big) + \big( \f{D} \ud{\Ps}(x) , {\accentset{.}{\star} \f{D} \ud{\Ps}} \big) + \big( \ud{\Ps} \ud{\Ps} , {\accentset{.}{\star} \ud{\Ps} \ud{\Ps}} \big) \Big\} \label{sa4}
\eeq
The resulting action for the fermionic part of the superconnection is second order in derivatives, not first order as in Dirac's action. Also, the last term in (\ref{sa4}) does not necessarily vanish, possibly giving four-fermion interactions. In components, the second-order fermion action is
\begin{eqnarray}
S_\Ps &=& \fr{V}{2} \int_M \big( \f{D} \ud{\Ps}(x) , {\accentset{.}{\star} \f{D} \ud{\Ps}} \big) \notag \\
&=& \fr{V}{2} \int_M \nf{d^M \! x} \, (D_i \accentset{.}{\Ps}^A) g^{ij} n_{AB} (D_j \ud{\Ps}^B) \notag \\
&=& \fr{V}{2} \int_M \nf{d^M \! x} \, (D_i \overline{\Ps}) (D^i \ud{\Ps}) \label{sa5}
\end{eqnarray}
in which $D_i \ud{\Ps}^A = \pa_i \ud{\Ps}^A + \lb C_i,\ud{\Ps} \rb^A$ is the covariant derivative, the metric, $g^{xy}$, along the fermionic directions has been used to define the fermion conjugate, $\accentset{.}{\Ps}$, such that, for example, $\accentset{.}{a} \ud{b}$ is a c-number, and the Lie algebra Killing form, $n_{AB}$, has been used to define the spinor conjugate, $\overline{\Ps}$. Although a second-order action for fermions is unusual, it is not incompatible with known physics, and the Standard Model can indeed be formulated with a second-order fermion action~\cite{Second3}. Even though the fields, $\ud{\Ps}$, in our action are essentially fermionic Klein-Gordon fields, they do transform under the spin connection, $\f{\om}$, in $\f{C}$ as spinors if they are valued in the spinorial part of a Lie algebra. Although it appears purely kinetic, our second-order fermion action (\ref{sa5}) includes an interaction with the frame-Higgs, $\f{e} \ph$, in $\f{C}$ that can give the fermions mass. If we assume a vacuum expectation value of the Higgs, $\ph_0 = \fr{\al}{2} \ga_4$, and that $\Ps$ is spinorial, then mass arises from
\beq
(D_i \overline{\Ps}) (D^i \ud{\Ps}) \sim (\overline{\Ps} C_i) (C^i \ud{\Ps}) \sim (\overline{\Ps} e_i \ph_0)(e^i \ph_0 \ud{\Ps})
= \overline{\Ps} e_i{}^\mu \fr{\al}{2} \ga_{\mu 4} \, e^{i\nu} \fr{\al}{2} \ga_{\nu 4} \ud{\Ps}
= \al^2 \, \overline{\Ps} \ud{\Ps} \label{fmass}
\eeq
with all fermions in $\ud{\Ps}$ having a bare mass of $\al$.

\section{Polynomial action}

In some approaches to quantum gravity it is desirable to formulate the theory with a polynomial action. If we wish to obtain our generalized Yang-Mills action (\ref{sa1}) from a polynomial action, we can introduce two auxiliary variables, $\nf{\cal B}$ and $\Ph$, and begin with the action~\cite{Smolin}
\beq
S = \int_{\tilde{G}} \Big\{ \lp \nf{\cal B} , \ff{\cal F} \rp + \fr{3}{4} \lp \nf{\cal B} , \Ph \nf{\cal B} \rp + \fr{1}{4} \lp \nf{\cal B} , \Ph \Ph \Ph \nf{\cal B} \rp \Big\} \label{bf}
\eeq
In this modified BF action, integrated over our $n$-dimensional deforming Lie group, the auxiliary variable $\nf{\cal B}(z)$ is a $Lie(G)$ valued $(n-2)$-form, and $\Ph$ is a linear operator taking $Lie(G)$ valued $(n-2)$-forms to $Lie(G)$ valued $2$-forms, and vice versa. Varying the action (\ref{bf}), the resulting equations are solved when $\Ph = \nf{\accentset{\cal G}{\star}}$ is the Hodge star operator, and $\nf{\cal B} = \nf{\accentset{\cal G}{\star}} \ff{\cal F}$, reproducing the equations of motion from our Yang-Mills action (\ref{sa1}).

\section{Regional fermions}

In a large Lie group, such as $F_4$ or $E_8$, having several regions, with a superposed Cartan connection (\ref{gCct}) that includes triality-related regional spin connections and frames, physical fermions in those regions can be described by Grassmann fields that are spinors with respect to the corresponding regional spin connection. Specifically, for a $Spin(4,4)$ (or similar) subgroup of the large group, regional fermion generators in the $8$-dimensional positive spinorial, negative spinorial, and vector representation spaces can be related by triality:
$$
\ud{\Ps}_{\mbox{\tiny I}} = \ud{\Ps}_{\mbox{\tiny I}}^\ch Q_\ch \;\;\;\;\;\;\;\;\;\;\;\;\;\; \ud{\Ps}_{\mbox{\tiny II}} = \ud{\Ps}_{\mbox{\tiny II}}^\ch Q^-_\ch = \ud{\Ps}_{\mbox{\tiny II}}^\ch T Q_\ch \;\;\;\;\;\;\;\;\;\;\;\;\;\; \ud{\Ps}_{\mbox{\tiny III}} =  \ud{\Ps}_{\mbox{\tiny III}}^\ch V_\ch = \ud{\Ps}_{\mbox{\tiny III}}^\ch T^2 Q_\ch
$$
Together with the $28$ generators of $spin(4,4)$, these $24$ fermionic generators (and their Lie brackets) constitute the Lie algebra of the $52$-dimensional split real exceptional Lie group $F_{4(4)}$. With no intersection among the regional fermionic generators, the $Lie(F_{4(4)})$ valued superconnection can be written as
\beq
\udf{G}(x) = \f{C}_{\mbox{\tiny I,II,III}} + \ud{\Psi}_{\mbox{\tiny I}} + \ud{\Psi}_{\mbox{\tiny II}} + \ud{\Psi}_{\mbox{\tiny III}} \label{superpsi}
\eeq
and the extended generalized Yang-Mills action (\ref{sa1}) integrated over the entire deforming Lie group, $\tilde{F}_{4(4)}$, becomes
\begin{eqnarray}
S &=& \fr{V}{2} \!\! \int_M  \!\! \big( \ff{F}(x) , \nf{\star F} \big)
+ \fr{V_{\mbox{\tiny T}}}{2}  \!\! \int_M  \!\! \big( \f{D}_{\mbox{\tiny I}} \ud{\Ps}_{\mbox{\tiny I}}(x) , \accentset{.}{\star} \f{D}_{\mbox{\tiny I}} \ud{\Ps}_{\mbox{\tiny I}} \big)
+ \fr{V_{\mbox{\tiny T}}}{2}  \!\! \int_M  \!\! \big( \f{D}_{\mbox{\tiny II}} \ud{\Ps}_{\mbox{\tiny II}}(x) , \accentset{.}{\star} \f{D}_{\mbox{\tiny II}} \ud{\Ps}_{\mbox{\tiny II}} \big) \notag \\
&+& \fr{V_{\mbox{\tiny T}}}{2}  \!\! \int_M \!\! \big( \f{D}_{\mbox{\tiny III}} \ud{\Ps}_{\mbox{\tiny III}}(x) , \accentset{.}{\star} \f{D}_{\mbox{\tiny III}} \ud{\Ps}_{\mbox{\tiny III}} \big)
+ \fr{V_4}{2}  \!\! \int_M  \!\! \big( \ud{\Ps}_{\mbox{\tiny I,II,III}} \ud{\Ps}_{\mbox{\tiny I,II,III}} , \accentset{.}{\star} \ud{\Ps}_{\mbox{\tiny I,II,III}} \ud{\Ps}_{\mbox{\tiny I,II,III}} \big) \notag
\end{eqnarray}
The first term involves the curvature of the superposed Cartan connection (\ref{gCct}), and reduces to a single integral over spacetime. The other terms, functionals of three triality-related sets of fermions, can be converted by triality automorphism to integrals involving three sets of fermions,
$$
\ud{\Ps}_{1} = \ud{\Ps}_{1}^\ch Q_\ch = \ud{\Ps}_{\mbox{\tiny I}} \;\;\;\;\;\;\;\;\;\;\;\;\;\; \ud{\Ps}_{2} = \ud{\Ps}_{2}^\ch Q_\ch = T^2 \ud{\Ps}_{\mbox{\tiny II}} \;\;\;\;\;\;\;\;\;\;\;\;\;\; \ud{\Ps}_{3} = \ud{\Ps}_{3}^\ch Q_\ch = T \ud{\Ps}_{\mbox{\tiny III}}
$$
each now using the same set of positive chiral spinor generators as the first. With this conversion, and because the Killing form is invariant under triality, the action becomes
\begin{eqnarray}
S &=& \fr{V}{2} \!\! \int_M \!\! \big( \ff{F}(x) , \nf{\star F} \big)
+ \fr{V_{\mbox{\tiny T}}}{2} \!\! \int_M \!\! \big( \f{D}_{1} \ud{\Ps}_{1}(x) , \accentset{.}{\star} \f{D}_{1} \ud{\Ps}_{1} \big)
+ \fr{V_{\mbox{\tiny T}}}{2} \!\! \int_M \!\! \big( \f{D}_{2} \ud{\Ps}_{2}(x) , \accentset{.}{\star} \f{D}_{2} \ud{\Ps}_{2} \big) \notag \\
&+& \fr{V_{\mbox{\tiny T}}}{2} \!\! \int_M \!\! \big( \f{D}_{3} \ud{\Ps}_{3}(x) , \accentset{.}{\star} \f{D}_{3} \ud{\Ps}_{3} \big)
+ \fr{V_4}{2} \!\! \int_M \!\! \big( \ud{\Ps}_{1,2,3} \ud{\Ps}_{1,2,3} , \accentset{...}{\star} \, \ud{\Ps}_{1,2,3} \ud{\Ps}_{1,2,3} \big) \label{faction}
\end{eqnarray}
in which the triality-transformed derivatives are, for example, $\f{D}_{2} \ud{\Ps}_{2} = \f{d} \ud{\Ps}_{2} + \lb \f{C}_2 , \ud{\Ps}_{2} \rb$, and the inter-generational Hodge, $\accentset{...}{\star}$, accounts for the triality relationship between $\ud{\Ps}_{1,2,3}$. Each fermion action term includes an interaction with the frame-Higgs in that region, as in (\ref{fmass}), generating equal masses for those fermions,
\beq
\fr{V_{\mbox{\tiny T}}}{2} \!\! \int_M \!\! \big( \f{D}_{2} \ud{\Ps}_{2}(x) , \accentset{.}{\star} \f{D}_{2} \ud{\Ps}_{2} \big)
= \fr{V_{\mbox{\tiny T}}}{2} \!\! \int_M \!\!\! \nf{d^M \! x} \, (D_{2i} \overline{\Ps}_2) (D_2^{i} \ud{\Ps}_2)
\sim \fr{V_{\mbox{\tiny T}}}{2} \!\! \int_M \!\!\! \nf{d^M \! x} \, (\al_2)^2 \, \overline{\Ps}_2 \ud{\Ps}_2 \label{3fmass}
\eeq

\section{Connections to physics}

The geometric framework of extended generalized Cartan geometry is sufficient to describe the structure and dynamics of gravitation and the Standard Model. The one field needed to describe this geometry is a superconnection valued in a large Lie algebra. To describe known physics the Lie algebra must include the structure of
\beq
spin(1,3) + \text{frame} \otimes \text{Higgs} + su(2)_L + su(1)_Y + su(3) \label{sma}
\eeq
acting on three fermion generations of $32$ complex or $64$ real spinor degrees of freedom each (or $60$ if neutrinos are Majorana).

It has been established that the algebraic structure of gravity and the Standard Model (\ref{sma}), with one generation of fermions, embeds in $spin(11,3)$ acting on a real chiral $64$ spinor representation space~\cite{Percacci, Lisi2}. However, to recover de Sitter spacetime, we require a Higgs direction with a spacelike signature, so we must use at least $spin(11,4)$. But $spin(11,4)$ does not have a suitable spinor representation, so we must turn to $spin(12,4)$, which does have a $128$-dimensional real chiral representation space that can accommodate one generation of fermions, as well as their mirrors. In this way it is possible to construct a non-simple Lie algebra accommodating the standard model with at least one generation~\cite{Douglas}. Even though this straightforward construction is possible, and can be extended to describe three generations of fermions, the appearance of mirror fermions is problematic and the model is not significantly simpler than the Standard Model it seeks to describe; also, the insertion of Yukawa coupling matrices between the Higgs and the generations, necessary to match known physics, needs to be put in by hand. Fortunately there is a superior possibility.

The algebra $spin(12,4) + S^+_{128}$, known to contain the algebra of gravity and the Standard Model with at least one generation of fermions, matches the Lie algebra of the largest simple quaternionic exceptional Lie group, $E_{8(-24)}$~\cite{Lisi2}. As well as the $spin(12,4) + S^+_{128}$ decomposition, there is another decomposition of the $E_{8(-24)}$ Lie algebra,
$$
E_{8(-24)} = spin(4,4) + spin(8) + S^+_8 \otimes S^+_8 + S^-_8 \otimes S^-_8 + V_8 \otimes V_8
$$
with three blocks of generators related by triality. These three blocks, spanned by $64$ generators each, accommodate the three generations of Standard Model fermions.

To match the Standard Model, we must presume that each generation of fermions is only accessible from one of three triality-related regions of $E_8$; only then do each of these regional fermion generations correctly match known fermion properties. Each generation will transform under a different triality-related regional spin connection, frame-Higgs, and set of gauge fields, consistent with the action (\ref{faction}) over the entire deforming Lie group. Because the different regional fermion generations transform under different $spin(1,3)$ Lorentz subalgebras of $E_8$, but with equal spin connection coefficients, all three generations of chiral fermions transform correctly, with no mirror fermions. Also, because the vacuum Higgs field may be different in each region, each generation of fermions may have a different bare mass. 

The entire algebra of gravitational and standard model bosons (\ref{sma}) cannot embed in $spin(4,4) + spin(8)$; however, we may embed all but the weak part of the algebra, implying that fermion weak eigenstates must be different than massive fermion eigenstates. More precisely, the $su(2)_L$ cannot fit in $spin(4,4) + spin(8)$ with the rest of (\ref{sma}), but must have at least two degrees of freedom, $W^+$ and $W^-$, in $V_8 \otimes V_8$, which will relate to others in $S^+_8 \otimes S^+_8$ and $S^-_8 \otimes S^-_8$ regions by triality. This fact has important consequences. Presuming that the weak bosons take the place of triality-related sterile neutrinos, the existing physical neutrinos must be Majorana. Also, the triality-related massive fermion states will not be weak eigenstates. The weak bosons will interact with mixed massive eigenstates, consistent with the CKM and PMNS matrices needed to rotate between mass and weak eigenstates. With all fields related by triality, as in (\ref{gCct}) and (\ref{superpsi}), the $E_8$ superconnection is
\begin{eqnarray}
\udf{G}(x) &=& \ha \f{\om}_{\mbox{\tiny I,II,III}} + \f{E}_{\mbox{\tiny I,II,III}} + \f{W}_{\mbox{\tiny I,II,III}} + \f{B}_{\mbox{\tiny I,II,III}} + \f{X}_{\mbox{\tiny I,II,III}} + \f{g} \notag \\
                 &  & + \lp \ud{\nu}_{eL} + \ud{e}_L + \ud{\bar{e}}_L + \ud{u}^{rgb}_L + \ud{\bar{u}}^{\, \bar{r}\bar{g}\bar{b}}_L + \ud{d}^{rgb}_L + \ud{\bar{d}}^{\, \bar{r}\bar{g}\bar{b}}_L  \rp \notag \\
                 &  & + \lp \ud{\nu}_{\mu L} + \ud{\mu}_L + \ud{\bar{\mu}}_L + \ud{c}^{rgb}_L + \ud{\bar{c}}^{\, \bar{r}\bar{g}\bar{b}}_L + \ud{s}^{rgb}_L + \ud{\bar{s}}^{\, \bar{r}\bar{g}\bar{b}}_L  \rp \notag \\
                 &  & + \lp \ud{\nu}_{\ta L} + \ud{\ta}_L + \ud{\bar{\ta}}_L + \ud{t}^{rgb}_L + \ud{\bar{t}}^{\, \bar{r}\bar{g}\bar{b}}_L + \ud{b}^{rgb}_L + \ud{\bar{b}}^{\, \bar{r}\bar{g}\bar{b}}_L  \rp \notag
\end{eqnarray}
with regional fermion generators related by triality; for example,
$$
\ud{e}_L = e^\chi_L \ud{Q}_\chi \;\;\;\;\;\;\;\;\;\;\;\;\;\;\;\; \ud{\mu}_L = \mu^\chi_L T \ud{Q}_\chi \;\;\;\;\;\;\;\;\;\;\;\;\;\;\;\; \ud{\ta}_L = \ta^\chi_L T^2 \ud{Q}_\chi
$$
The triality-related boson generators in the three regions act on the corresponding fermions in agreement with their familiar Standard Model spins and charges. With all known particles matched to algebraic elements of $E_8$, there is one new gauge boson (and its anti-particle) remaining, with electric charge $\pm \fr{4}{3}$ and weak charge $\pm \ha$, having three colors. If this $X$ boson exists, it could bind with two up quarks in a massive, electrically neutral, spin one composite particle. This new $X$ boson is one of the $SO(10)$ $X$ bosons, and would allow proton decay via the channel $u u d \to u u X^{-\fr{4}{3}} \, \bar{e} \to u \bar{u} \, \bar{e}$.


\pagebreak
\begin{figure}[h!]
\centering
\includegraphics[width=1.0\textwidth]{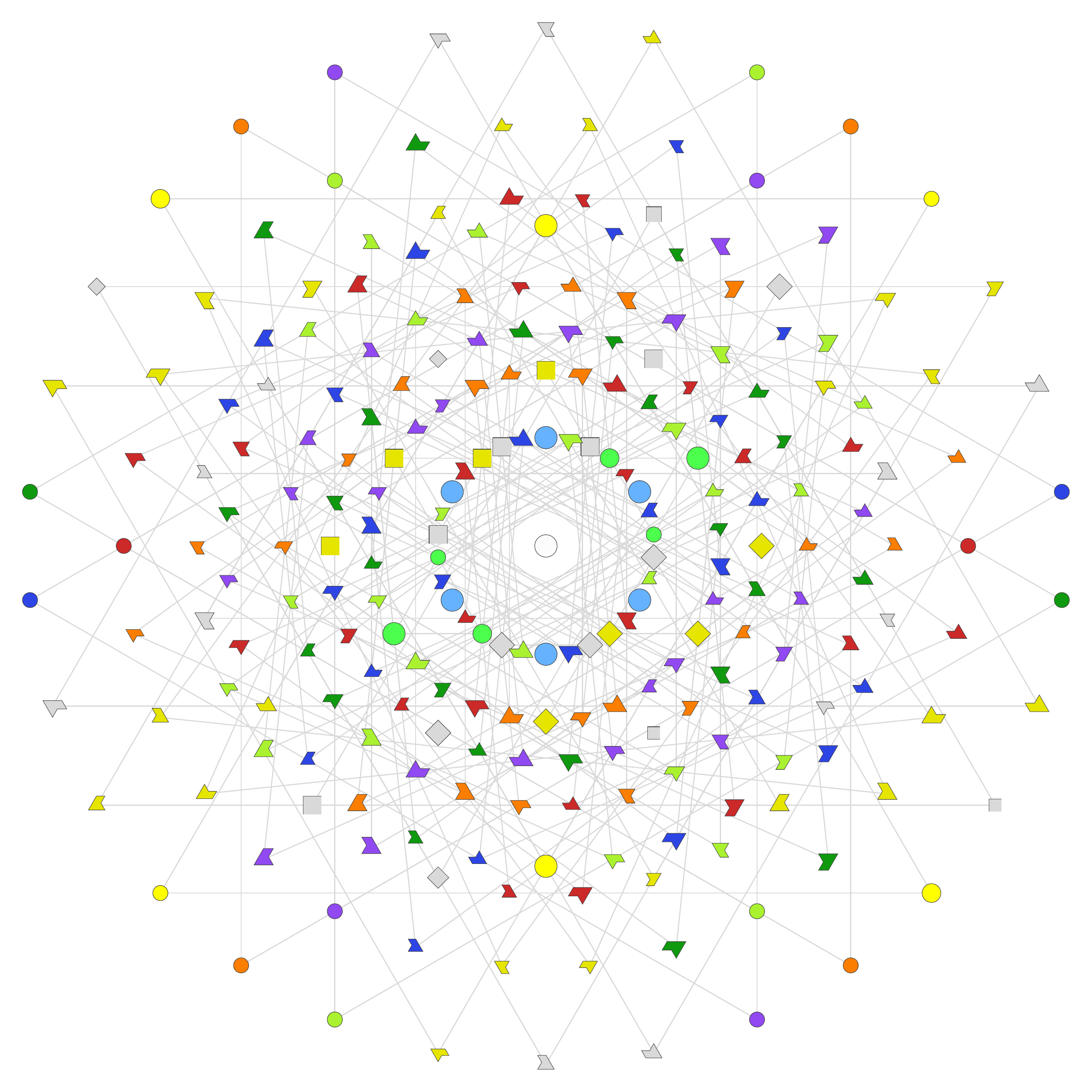}
\caption{The $E_8$ root system, with three generations of particles related by triality. These particle states are meant to be suggestive rather than definitive. The detailed assignments of elementary particle states to $E_8$ roots, views of other rotations, and other unification models, are available at the Elementary Particle Explorer: \texttt{\href{http://deferentialgeometry.org/epe/}{http://deferentialgeometry.org/epe/}}}
\end{figure}
\pagebreak

\section{Summary}
\label{sec:summary}

We propose that our universe is an excited Lie group, described by a superconnection valued in a large Lie algebra, with regional fermion generations related by triality automorphisms, and dynamics governed by a generalized Yang-Mills action.

The local geometry of an unexcited Lie group, $G'$, is described by its natural Maurer-Cartan form (\ref{MC}),
\beq
\f{\Th} = g^- \f{d} g = \f{\xi}^A T_A \label{smcf}
\eeq
understood as both a connection and frame, which, along with the Lie algebra's Killing form (\ref{kf}), determines a metric on the Lie group manifold. For the $G'=Spin(1,4)$ Lie group, a four-dimensional subgroup, $S \subset G'$, can be chosen such that its Maurer-Cartan form is
$$
\f{\Th}^{\mbox{\tiny S}} = \ha \f{\om}^{\mbox{\tiny S}} + \f{E}^{\mbox{\tiny S}} \;\;\;\;
\f{\om}^{\mbox{\tiny S}} = - \f{ds}^\pi \al e^{\al t} \ga_{0 \pi} \;\;\;\;
\f{E}^{\mbox{\tiny S}} = \f{e}^{\mbox{\tiny S}} \ph_{\mbox{\tiny S}} \;\;\;\;
\f{e}^{\mbox{\tiny S}} =  \f{dt} \ga_0 + \f{d s^\pi} e^{\al t} \ga_\pi \;\;\;\;
\ph_{\mbox{\tiny S}} = \ph_0^4 \ga_4 = \fr{\al}{2} \ga_4
$$
with $\f{\om}^{\mbox{\tiny S}}$ and $\f{e}^{\mbox{\tiny S}}$ the spin connection (\ref{dSw}) and gravitational frame (\ref{dSe}) for de Sitter spacetime, and $\ph_0^4 = \fr{\al}{2}$ a Higgs vev, with expansion parameter $\al$. This de Sitter subgroup, $S$, is also a coset representative of $G'/H$, with $H=Spin(1,3)$ a chosen Lorentz subgroup of $Spin(1,4)$. Deformations of $G'$ maintaining the structure of the gauge group, $H$, are described by allowing this Maurer-Cartan form, $\f{\Th}^{\mbox{\tiny S}}$, to vary, becoming a Cartan connection (\ref{ccon}),
\beq
\f{C}(x) = \ha \f{\om} + \f{e} \ph_{\mbox{\tiny S}} \label{sccon}
\eeq
valued in $Lie(G')=spin(1,4)$, on an arbitrary four-dimensional spacetime, $M$, modeled on $S$, with nontrivial curvature. The Maurer-Cartan form (\ref{smcf}), $\f{\Th}$, over $G'$ varies accordingly and becomes the Ehresmann-Cartan connection form (\ref{eCc}), $\f{\cal C}$, over the deforming Lie group manifold, $\tilde{G'}$. The Cartan connection (\ref{sccon}) is the pullback of $\f{\cal C}$ onto embedded spacetime, $M$, representing a sheaf of gauge-related spacetimes embedded in $\tilde{G'}$.

For a larger Lie group, $G$, containing $G'$ as a subgroup, some excitations of $G$ may be described by a $Lie(G)$ valued generalized Cartan connection (\ref{gCc}),
$$
\f{C}(x) = \ha \f{\om} + \f{E} + \f{A}
$$
defined on four-dimensional spacetime, $M$, embedded in $\tilde{G}$, modeled on $S$ embedded in $G$. More general excitations of $G$ are described by a new kind of superconnection (\ref{sc}), 
$$
\udf{G}(x) = \f{C} + \ud{\Ps}  \;\;\;\;\;\;\;\;\;\;
\ud{\Ps}(x) = \ud{\Ps}^\ch Q_\ch = \Ps^\ch \ud{Q}_\ch \;\;\;\;\;\;\;\;\;\;
\ud{Q}_\ch = \ud{\xi}^\ch Q_\ch \;\; \mbox{(no sum)}
$$
with $\f{C}$ and $\ud{\Ps}$ valued in complementary parts of $Lie(G)$, and the fermionic part (\ref{ferm}), $\ud{\Ps}$, understood as a field of $1$-forms in $\tilde{G}$ defined on and orthogonal to embedded spacetime, $M$, with super-generators, $\ud{Q}_\ch$, defined using vertical components, $\ud{\xi}^\ch \sim \f{\xi}^\ch$, of the Maurer-Cartan form (\ref{smcf}). 

Different, possibly intersecting Lie group regions may be related by a finite group, $\Ga$, of Lie group automorphisms. Regions can be covered by Lorentz and gauge transformations of spacetimes embedded in each region. For a Lie group such as $E_{8(-24)}$, having a $Spin(1,4)$ (or similar) subgroup and supporting triality automorphisms, excitations may be described by a superposed superconnection (\ref{superpsi}),
\beq
\udf{G}(x) = \f{C} + \ud{\Ps} = \f{C}_{\mbox{\tiny I}} + \f{C}_{\mbox{\tiny II}} + \f{C}_{\mbox{\tiny III}} + \ud{\Ps}_{\mbox{\tiny I}} + \ud{\Ps}_{\mbox{\tiny II}} + \ud{\Ps}_{\mbox{\tiny III}} \label{ssuperpsi}
\eeq
with regional bosons, $\f{C}_{\mbox{\tiny I,II,III}}$, and fermions, $\ud{\Ps}_{\mbox{\tiny I,II,III}}$, having generators related by triality,
$$
\ud{\Ps}_{\mbox{\tiny I}} = \Ps_{1}^\ch \ud{Q}_\ch \;\;\;\;\;\;\;\;\;\;\;\;\;\;\;\; \ud{\Ps}_{\mbox{\tiny II}} = \Ps_{2}^\ch T \ud{Q}_\ch \;\;\;\;\;\;\;\;\;\;\;\;\;\;\;\; \ud{\Ps}_{\mbox{\tiny III}} = \Ps_{3}^\ch T^2 \ud{Q}_\ch
$$
These regional fermions transform as spinors under different $Spin(1,3)$ subgroups, corresponding to different embedded spacetimes, with regional spin connections, $\f{\om}_{\mbox{\tiny I,II,III}}$, related by triality. Physical spacetime is a superposition of spacetimes in these three triality-related regions.

The motion of an excited Lie group is described by an extended generalized Yang-Mills action (\ref{sa1}) integrated over the entire deforming Lie group manifold, $\tilde{G}$, which reduces via symmetry to an action over spacetime (\ref{sa3}),
\beq
S = \fr{1}{2} \int_{\tilde{G}} \lp \ff{\cal F}(z) , \nf{\accentset{\cal G}{\star} \cal F} \rp = \fr{V}{2} \int_M \lp \udff{F}(x) , \nf{\accentset{.}{\star} F} \rp \label{action}
\eeq
in which the supercurvature (\ref{scurv}), is
$$
\udff{F}(x) = \f{d} \, \udf{G}(x) + \ha [ \udf{G}, \udf{G} ]
= \lp \f{d} \f{C} + \f{C} \f{C} \rp + \lp \f{d} \ud{\Ps} + [\f{C}, \ud{\Ps}] \rp + \lp \ud{\Ps} \ud{\Ps} \rp
= \ff{F} + \f{D} \ud{\Ps} + \ud{\Ps} \ud{\Ps}
$$
with the curvature of the generalized Cartan connection (\ref{gCccurv}) equal to
$$
\ff{F} = \f{d} \f{C} + \f{C} \f{C} = \lp \ha \ff{R} + \f{E} \f{E} \rp + \f{D} \f{E} + \ff{F}_A
$$
and the super-Hodge star, $\accentset{.}{\star}$, naturally derived from the Killing form and the frame part, $\f{E}$, of $\f{C}$, producing fermion conjugation. Using this supercurvature, and the triality-related fermion generations (\ref{ssuperpsi}), the action (\ref{action}) becomes (\ref{faction}),
\begin{eqnarray*}
S &=& \fr{V}{2} \!\! \int_M \! \Big\{ \big( \ff{F}(x) , \nf{\star F} \big) + \big( \f{D} \ud{\Ps}(x) , {\accentset{.}{\star} \f{D} \ud{\Ps}} \big) + \big( \ud{\Ps} \ud{\Ps} , {\accentset{.}{\star} \ud{\Ps} \ud{\Ps}} \big) \Big\} \\
&=& \fr{V}{2} \!\! \int_M \! \Big\{ \big( \ff{R} , \star \f{E} \f{E} \big) + \big( \f{E} \f{E} , \star \f{E} \f{E} \big) + \fr{1}{4} \big( \ff{R} , \star \ff{R} \big)
+ \big( \f{D} \f{E}  , \star \f{D} \f{E} \big) + \big( \ff{F}_A  , \star \ff{F}_A \big) \Big\} \\
& & + \sum_{\al=1}^3 \fr{V_{\mbox{\tiny T}}}{2} \!\! \int_M \!\! \nf{d^M \! x} \, (D^\al_i \overline{\Ps}_\al) (D^{\al i} \ud{\Ps}_\al) + O(\Ps^4)
\end{eqnarray*}
This action, obtained purely from a generalized Yang-Mills functional, matches the action of gravity and the three-generation Standard Model, with a few irregularities: The usual gravitational action is amended by a Kretschmann scalar term, $\ff{R}  \star \ff{R}$. Also, the fermion action is second-order in derivatives, and possibly includes a four fermion interaction term. And, in this model, the fermions of each generation naturally interact with the Higgs vev and obtain identical bare masses.

When all fields and particles of General Relativity and the Standard Model, including three generations of fermions, are described in this manner as excitations of the largest simple quaternionic Lie group, $E_{8(-24)}$, having $248$ dimensions, there is one new, colored gauge boson and its anti-particle remaining, $X$, with electric charge $\pm \fr{4}{3}$.

\section{Discussion}

In this work we have attempted to elucidate a minimal geometric model, Lie Group Cosmology (LGC), capable of describing all known particles, fields, and dynamics of General Relativity and the Standard Model as excitations of a single Lie group governed by an extended Yang-Mills action. Our spacetime's vacuum state is represented as a four-dimensional de Sitter subgroup of the Lie group, with geometry described by the Lie group's natural Maurer-Cartan connection. When this Lie group is excited, or deforms, this natural connection varies away from its state of zero curvature, becoming a new kind of Cartan connection. One part of this new, Lie algebra valued connection includes four spacetime basis generators, providing the gravitational frame for the embedded spacetime. Another part of the new connection is the spin connection for this frame, while other parts include gauge and scalar fields over spacetime. This new connection is further generalized to a natural superconnection, having components that are $1$-forms orthogonal to embedded spacetime, which can describe physical fermions. In addition to Lorentz and conventional gauge transformations, other automorphisms, such as triality, can relate spacetimes in different regions of the deforming Lie group. With fermions described as spinors on spacetimes embedded in different, triality-related regions, there are naturally three generations of massive, chiral fermions. An extended Yang-Mills action for the superconnection, integrated over the entire deforming Lie group manifold, reduces to a Yang-Mills action for bosons and a second-order action for the regional fermions. When gravitational and Standard Model fields, with three generations of fermions, are matched in this way to the $248$-dimensional exceptional Lie group, $E_8$, there is a new $X$ boson predicted.

Lie Group Cosmology is a general, minimal, natural geometric framework, capable of describing gravity, the Standard Model, and some structures beyond the Standard Model. Although LGC could stand on its own as a theoretical framework, it solves several problems and mysteries raised in ``An Exceptionally Simple Theory of Everything''~\cite{Lisi1}, using a more natural structure. With a single deforming Lie group, rather than a principal bundle, as the fundamental structure, described by a new kind of Cartan superconnection, many issues are resolved in a succinct geometric model. The spacetime frame is not mysteriously selected from among Lie group generators, but comes from the generators along spacetime embedded in the deforming Lie group, and the deforming Lie group's Hodge star reduces to the spacetime Hodge. In this theory, physical fermions are understood naturally as excitations of the superconnection orthogonal to spacetime. With three sheaves of spacetime embedded in different regions, there are three sets of regional chiral fermions, with correct quantum numbers, corresponding to three generations of Standard Model fermions, and no mirror fermions. Also, the LGC action is succinct: an extended Yang-Mills action governing Lie group excitations described by the superconnection, capable of prescribing the dynamics and interactions of gravity and the Standard Model.

Although it is surprising that a model as simple as LGC works as well as it does, there are several irregularities and problems. Most importantly, the precise mechanism by which the CKM and PMNS matrices, and particle masses, emerge from the theory is not yet clear. For this reason, LGC and its correct application to the Standard Model, via E8 Theory, cannot yet be considered complete. It may be possible that some choice of split, quaternionic, or complex form of $E_8$, along with the correct LGC model and a more complex triality automorphism, will lead to a complete picture---but that awaits further work. Also, the LGC action, resulting in an unusual action for fermions and an extra gravitational term, may prove untenable. In addition, in unified theories, new particles predicted might not be consistent with observation. There are also conceptual issues with LGC. ``Symmetry keeping'' is central to the theory, but there is no explanation as to why some specific symmetries would be preserved and not others. There is no good reason why only four-dimensional subspaces of the original Lie group, or their related copies in different regions, would deform. One philosophical justification could be that the geometry and topology of four-dimensional manifolds is maximally rich~\cite{Scorpan}, and that the representations of the $E_8$ Lie group are the most numerous, but this is not completely satisfying. Also, there is no good reason, other than the spin-statistics requirement, why the bosonic and fermionic parts of the superconnection must be valued in complementary parts of the Lie algebra. It is possible that this restriction could be relaxed, allowing the existence of BRST ghosts or other particles, or that there may be a natural reason for the restriction that is not yet clear. A better understanding is needed. Also, in brazenly attempting to advance our understanding of fundamental physics, it is likely we have made mistakes. It is possible that mistakes have been made in the mathematical formulation of the theory, or in its application to describing our world. Whenever new territory is explored, there are missteps, and we can only hope these mistakes can be corrected with further knowledge.

Despite its deficiencies, LGC does provide an unusually successful model for fundamental physics. It also motivates several highly speculative areas of investigation. Since excitations of a noncompact Lie group play the dominant role in the theory, progress might be made by furthering Harish-Chandra's program of harmonic analysis and representation theory, extending parallels between the Peter-Weyl theorem, representation theory, and Quantum Field Theory, for noncompact Lie groups. Also, on a purely philosophical level, the question arises of whether there could be any importance to the identity element. The minimal fundamental structure in LGC is a deforming torsor, but the theory works just as well with a deforming Lie group, including a distinguished identity element. This motivates the crazy idea that the group identity element could play the role, philosophically, of the ontological identity---physically, the existence of the observer at a distinguished ``present'' spatial position and temporal moment in spacetime. This highly speculative idea might lead to new insights on the quantum measurement problem. Also, more conservatively, since the fundamental field in LGC is a connection, the theory is compatible with some Loop Quantum Gravity approaches to quantization. Lie Group Cosmology also suggests that our universe is fundamentally de Sitter---infinite in spatial and both temporal directions. Physically, this means our universe existed forever, before the big bang, and has been forever cooling and accelerating in its expansion, and will continue this expansion, approaching perfect emptiness and symmetry.

The geometric picture of Lie group unification is concise. The vacuum state of the total space of physics, consisting of base spacetime and fields valued in various fibers, is a single Lie group manifold with its natural connection. Deformations of this Lie group and its de Sitter submanifolds, described by variations of the connection away from zero curvature, correspond to excitations of fields away from this vacuum state. Spacetime is the part of the Lie group that is deforming, while fibers are subspaces maintaining their structure. The $1$-form, scalar, and fermionic parts of the superconnection are physical fields over deforming four-dimensional spacetime, with dynamics governed by an extended Yang-Mills action. In this way, we find ourselves in an approximately de Sitter spacetime alive with excitations, described by variations of Lie algebra valued fields. The existence of our universe, including the structure of spacetime and all fields, may be understood as deformations of a single Lie group. The reason we see Lie groups and their representations everywhere we look is because we are inside a deforming Lie group, looking out.



\end{document}